\pdfoutput=1
\documentclass[12pt]{article}
\usepackage[utf8]{inputenc}
\usepackage{tabularx}
\usepackage{longtable}

\usepackage[left=2cm,right=2cm,top=2cm,bottom=2cm]{geometry}   
\usepackage[small]{caption}
\usepackage{relsize}
\usepackage[table,svgnames,dvipsnames]{xcolor}
\usepackage{enumitem}
\usepackage{bbold}
\usepackage[toc,page]{appendix}
\usepackage{listings}   
\usepackage[pdftex]{graphicx}
\usepackage{latexsym,amsmath,amsfonts,amssymb}
\usepackage[numbers,sort&compress]{natbib}
\usepackage{url}
\usepackage[colorlinks=true
  ,urlcolor=blue
  ,anchorcolor=blue
  ,citecolor=blue
  ,filecolor=blue
  ,menucolor=blue
  ,linktocpage=true
  ,pdfproducer=medialab
  ,pdfa=true
  ,linktoc=all
]{hyperref}

\usepackage{amssymb}                                
\usepackage{amsmath}                    			
\usepackage{mathtools}                  			
\usepackage{latexsym}                   			
\usepackage{bbold}                      			
\usepackage{braket}                     			
\usepackage{graphicx}               		        
\usepackage{subcaption}
\usepackage{wrapfig}                    			
\usepackage{sidecap}                    			
\usepackage{physics}                    			
\usepackage{color}                      			
\usepackage[dvipsnames]{xcolor}         		    
\usepackage{soul}                       			
\usepackage{array}                      			
\usepackage{enumitem}                   			
\usepackage{tensor}                     			
\usepackage{cancel}                     			
\usepackage{comment}                    			
\usepackage{tikz}
\usepackage{bbm}
\usepackage{float}
\usepackage{pdfscreen}                              
\usepackage[display]{texpower}                      
\usepackage{pifont}
\usepackage[thinlines]{easytable}
\usepackage{tikz}
\usetikzlibrary{decorations.pathreplacing}
\usetikzlibrary{decorations.pathmorphing}
\usepackage{indentfirst}                            
\usepackage{csquotes}                               
\MakeOuterQuote{"}                                  

\numberwithin{equation}{section}

\begin{document}

\begin{center}
\centerline{\Large {\bf Floquet SYK wormholes}}

\vspace{8mm}
\hypersetup{linkcolor=black}
\renewcommand\thefootnote{\mbox{$\fnsymbol{footnote}$}}
Mart\'i Berenguer\footnote{marti.berenguer.mimo@usc.es},
Anshuman Dey\footnote{anshuman.dey@usc.es},
Javier Mas\footnote{javier.mas@usc.es},
Juan Santos-Su\'arez\footnote{juansantos.suarez@usc.es}
and  Alfonso V. Ramallo\footnote{alfonso.ramallo@usc.es}

\vspace{4mm}

{\small \sl Departamento de  F\'isica de Part\'iculas} \\
{\small \sl Universidade de Santiago de Compostela} \\
{\small \sl and} \\
{\small \sl Instituto Galego de F\'isica de Altas Enerx\'ias (IGFAE)} \\
{\small \sl E-15782 Santiago de Compostela, Spain} 
\vskip 0.2cm

\end{center}

\vspace{8mm}
\numberwithin{equation}{section}
\setcounter{footnote}{0}
\renewcommand\thefootnote{\mbox{\arabic{footnote}}}

\begin{abstract}

We study the non-equilibrium dynamics of two coupled SYK models, conjectured to be holographically dual to an eternal traversable wormhole in AdS$_2$. We consider different periodic drivings of the parameters of the system. We analyze the energy flows in the wormhole and black hole phases of the model as a function of the driving frequency. Our numerical results show a series of resonant frequencies in which the energy absorption and heating are enhanced significantly and the transmission coefficients drop, signalling a closure of the wormhole. These frequencies correspond to part of the conformal tower of states and to the boundary graviton of the dual gravitational theory. Furthermore, we provide evidence supporting the existence of a hot wormhole phase  between the black hole and wormhole phases. When driving the strength of the separate SYK terms we find that the transmission can be enhanced by suitably tuning the driving.  
\end{abstract}

\newpage

\hypersetup{linkcolor=black}
\tableofcontents

\hypersetup{linkcolor=blue}
\section{Introduction}
The Sachdev-Ye-Kitaev (SYK) model \cite{Sachdev_1993,Kitaevtalk1,KitaevTalk2} is a quantum mechanical model involving $N$ Majorana fermions with random all-to-all quartic interactions. In the large $N$ limit, this model admits a dual  description at low-energy in terms of Jackiw-Teitelboim (JT) gravity on AdS$_2$, providing a neat example of the holographic field theory/gravity duality (see \cite{Sarosi_2018,Trunin_2021,Chowdhury_2022} for reviews). The thermal AdS$_2$ space has two boundaries that are causally disconnected. As shown in \cite{Gao_2017}, by adding a suitable double trace deformation, these boundaries can be brought instantaneously into causal contact. In \cite{MaldaQi}, Maldacena and Qi proposed an extended dual version of this mechanism by coupling together two SYK models. For small coupling $\mu$, the ground state of this model is dual to an eternal traversable wormhole in the global AdS$_2$ spacetime, where the two copies of the SYK model sit at its boundaries.

The SYK model and its generalizations have been widely studied both in the high-energy and condensed matter contexts. In the latter, periodically driven many-body systems provide a remarkable laboratory for exploring out-of-equilibrium physics. Not only can one understand the long-time behavior of the system, but eventually, discover methods to generate new states exclusive to such non-equilibrium regimes, a.k.a. Floquet engineering \cite{Bukov_2015,rudner2020floquet,Behrends_2022}. The present paper investigates the response of the two coupled SYK model \cite{MaldaQi} to periodic modulations.

Our original motivation to study this problem was twofold. On one hand, we wanted to learn how, within this model, the degrees of freedom of the wormhole couple to external perturbations that are periodic in time. This may give some insight into their quantum internal makeup. In fact, we found an intriguing selection rule in the case we studied more thoroughly. On the other hand, we wondered whether periodic drivings could help to understand, and eventually improve, the wormhole-inspired teleportation protocols that have recently captured the interest of the quantum computing community \cite{Brown_2023, Nezami_2023}. Remarkably, we have found evidence supporting that, when tuned appropriately, periodic drivings can enhance the transmission signal.

The model in \cite{MaldaQi} exhibits two phases, whose properties consistently match with the dual side of the geometry of a traversable wormhole, and two disconnected black holes, respectively. It is believed that these two phases are connected through a canonically unstable, but microcanonically stable phase, known as the "hot wormhole" \cite{MaldaQi,MaldaMilekhin}. This unstable phase is also present in the case the Majorana fermions are replaced by complex fermions \cite{GarciaGarcia_2021}. Considerable effort has been dedicated to examining this model and variants thereof. To our knowledge, only a handful of studies involve time-dependent dynamics.
In \cite{MaldaMilekhin}, the authors address the dynamical formation of the wormhole.
Other works instead deal with the time evolution after a sudden quench \cite{Eberlein_2017,Bhattacharya_2019,Larzul_2022, Zhang_2021}.

The generic late-time expectation regarding the response to periodic drivings is a net energy injection and a subsequent heating towards an infinite temperature state. In some cases, this process can be strongly suppressed, and systems can enter a pre-thermal regime with an effective temperature which is stable over long times. In the context of the complex SYK, this has been shown to be the case in the Fermi liquid phase (FL) due to the presence of quasi-particles, to be contrasted with the rapid thermalization in the non-Fermi liquid phase (NFL) \cite{Kuhlenkamp_2020}.

In this paper, we conduct drivings akin to those described in \cite{Kuhlenkamp_2020}, this time for the two-coupled SYK model introduced in \cite{MaldaQi}. Specifically, we drive the relevant coupling, $\mu$, sinusoidally in time. The analog of the NFL vs FL phase manifests here in the wormhole and black hole phases respectively, where the former is gapped while the latter is not. We also consider drivings of the interaction strengths $J_{L,R}$ of each SYK independently. These last are less suitable to a physical implementation, as they imply a time variation of the overall variance of the coupling distribution function.

For the driving in $\mu$, the overall picture looks consistent with the expectations. In the wormhole phase, however, a significant heating amplification appears when driving at specific subsets of the eigenfrequencies of the undriven system. While the resonant behavior seems intuitive, understanding why the driving couples only to half of the spectrum requires careful consideration. Of particular interest is the second resonance, which appears to correspond to exciting a boundary graviton mode.

Also, by fine-tuning the duration of the driving, we are able to obtain a series of asymptotic equilibrium states that are compatible with the hot wormhole, the canonically unstable phase that joins the two stable phases, but that had not been found in an equilibrium analysis of the system.

The present paper is organized as follows. We will set the stage in section \ref{sec:eqform} by reviewing the non-equilibrium formalism of the model as well as the observables that will be used to monitor the effects of the driving. The next section, \ref{sec:results}, shows the results of our simulations. We also provide details of the numerical tools that have been used in Appendix \ref{App:initialconds}. We finally address the Schwarzian calculation in section \ref{sec:schwarz} and then wrap up with a discussion and future directions.

\section{Formalism}
\label{sec:eqform}

In this section we will first define the model and summarise the equations of motion in the large $N$ limit. Then we will review the two phases of the model and collect the main observables of interest. These include the energy and effective temperature. 

\subsection{Equations of motion}

The two-coupled SYK model starts by considering
$2N$ Majorana fermions $\chi_a^i$ with $i=1,...,N$ and $a = L,R$, satisfying the usual anticommutation relations $\left\{\chi_a^i,\chi_b^j\right\}=\delta^{ij}\delta_{ab}$. The relevant dynamics is contained in the  following time dependent Hamiltonian 
\begin{equation}
H(t)=\sum_{a=L,R}\frac{1}{4!}\sum_{ijkl}f_a(t)J_{ijkl}\chi_a^i\chi_a^j\chi_a^k\chi_a^l+i\mu(t)\sum_j \chi_L^j\chi_R^j ~,
\label{eq:tpham}
\end{equation}
where $J_{ijkl}$ are real constants drawn from a gaussian distribution with mean and variance given by
\begin{equation}
\overline{J_{ijkl}}=0~,~~~~~\overline{J_{ijkl}^2}=\frac{3! J^2}{N^3} ~.
\label{eq:Jmeanvar}
\end{equation}
Setting $\mu(t) = \mu$ to a constant value and $f_{L/R}(t)\rightarrow 1$ recovers the model dual to a traversable wormhole in \cite{MaldaQi}. 

\begin{figure}
    \centering
    \includegraphics[width=0.9\linewidth]{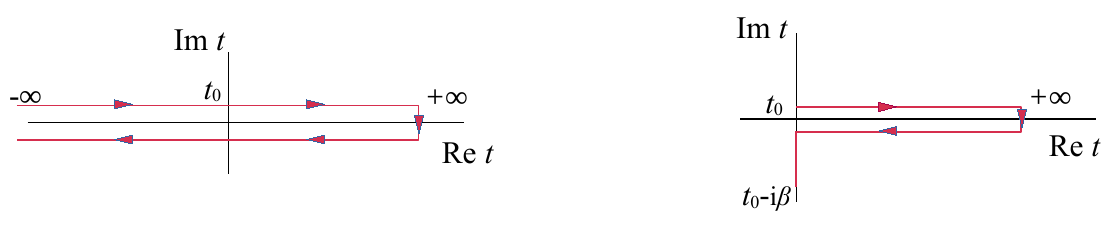}

    \caption{Contours used in the numerical integration. At $t=t_0$,  the drivings are turned on. Prior to that, the system is at equilibrium. When using the standard predictor-corrector method the initial condition must be explicitly provided (left). When using NESSi package, the L-shaped contour (right) calculates the initial thermal equilibrium state directly in the Matsubara formalism.  }
    \label{fig:Keldyshcontour}
\end{figure}

In order to study the non-equilibrium dynamics of the model proposed above, the Schwinger-Keldysh formalism is instrumental. In this formalism, observables are computed by evolving the system forwards and backwards in time along the  Keldysh contour ${\cal C}$ (see Fig. \ref{fig:Keldyshcontour}). The action is therefore
\begin{equation}
    S=\int_\mathcal{C}dt\Biggl[\frac{i}{2}\sum_{a=L,R}\sum_j\chi_a^j\partial_t\chi_a^j-\frac{1}{4!}\sum_{a=L,R}\sum_{i,j,k,l}f_a(t) J_{ijkl}~\chi_a^i\chi_a^j\chi_a^k\chi_a^l-\frac{i\mu(t)}{2}\sum_j\left(\chi_L^j\chi_R^j-\chi_R^j\chi_L^j\right)\Biggr]~.
\end{equation}

After performing the disorder average the partition function can be written in terms of an effective action $S_{eff}[G,\Sigma]$, whose arguments are the contour-ordered Green's functions
\begin{equation}
    i G_{ab}(t_1,t_2)=\frac{1}{N}\sum_j\langle \mathcal{T}_\mathcal{C}\chi_a^j(t_1)\chi_b^j(t_2)\rangle
    \label{eq:contourGF}
\end{equation}
and the associated self energy  $\Sigma(t_1,t_2)$. That is
\begin{equation}
\overline{Z}=\int\mathcal{D}G_{ab}\mathcal{D}\Sigma_{ab}\exp\left[\frac{iS_{eff}[G,\Sigma]}{N}\right]
\end{equation}
with
\begin{align}
\begin{split}
   \frac{iS_{eff}[G,\Sigma]}{N}=&\frac{1}{2}\log\det \left(-i\left[G_0^{-1}\right]_{ab}(t_1,t_2)-\mu_{ab}(t_1)\delta(t_1-t_2)+i\Sigma_{ab}(t_1,t_2)\right)\\
    &-\frac{1}{2}\int_\mathcal{C}dt_1dt_2\sum_{a,b}\Big(\Sigma_{ab}(t_1,t_2)G_{ab}(t_1,t_2)+\frac{J_L(t)J_R(t)}{4}G_{ab}(t_1,t_2)^4\Big)~,
\end{split}
\end{align}
where $J_a(t)\equiv J f_a(t)$ and
\begin{equation}
    \mu_{ab}(t)=\begin{pmatrix}
0 & \mu(t) \\
-\mu(t) & 0
\end{pmatrix}~.
\end{equation}

The large $N$ saddle point equations of motion are $\delta S_{eff}/\delta G_{ab}=0$ and $\delta S_{eff}/\delta 
\Sigma_{ab}
=0$.
In terms of the greater and lesser components, defined as follows
\begin{align}
\begin{split}
    G_{ab}^>(t_1,t_2)&=-\frac{i}{N}\sum_j\langle\chi_a^j(t_1)\chi_b^j(t_2)\rangle~,\\
    G_{ab}^<(t_1,t_2)&=-\frac{i}{N}\sum_j\langle\chi_b^j(t_2)\chi_a^j(t_1)\rangle~,
\end{split}
\end{align}
we can obtain the Kadanoff-Baym equations for the two-point functions $G_{ab}^>(t_1,t_2)$ (see  Appendix \ref{App:KBequations} for details)
\begin{eqnarray}
    i\partial_{t_1}G_{ab}^>(t_1,t_2)&=&i\mu_{ac}(t_1)G_{cb}^>(t_1,t_2)+\int_{-\infty}^{\infty}dt~ \Sigma_{ac}^R(t_1,t)G_{cb}^>(t,t_2)+\int_{-\infty}^{\infty}dt~ \Sigma_{ac}^>(t_1,t)G_{cb}^A(t,t_2)~,\nonumber \\
    -i\partial_{t_2}G_{ab}^>(t_1,t_2)&=& i G_{ac}^>(t_1,t_2)\mu_{cb}(t_2)+\int_{-\infty}^{\infty}dt~ G_{ac}^R(t_1,t)\Sigma_{cb}^>(t,t_2)+\int_{-\infty}^{\infty}dt ~G_{ac}^>(t_1,t)\Sigma_{cb}^A(t,t_2)~.
\nonumber \\
&& \label{eq:KBeqs}
\label{eq:Sigmarealt}
\end{eqnarray}
For the self-energies we find 
\begin{equation}
    \Sigma^>_{ab}(t_1,t_2) =-J_a(t)J_b(t)G^>_{ab}(t_1,t_2)^3~.
\end{equation}

Following the literature \cite{Haldar_2020,Bhattacharya_2019,Eberlein_2017,Kuhlenkamp_2020,Larzul_2022,MaldaMilekhin,Zhang_2021}, the prevailing approach to integrate these equations numerically involves a two-times grid and employs a predictor-corrector method. The reason a two-times grid is needed is because  out of equilibrium the Green's functions depend in general on two times, instead of  only on time differences, as it happens in equilibrium. This makes the non-equilibrium  numerical solving much more resource consuming than the equilibrium one.

\begin{figure}[h!]
    \centering
    \includegraphics[width=0.48\linewidth]{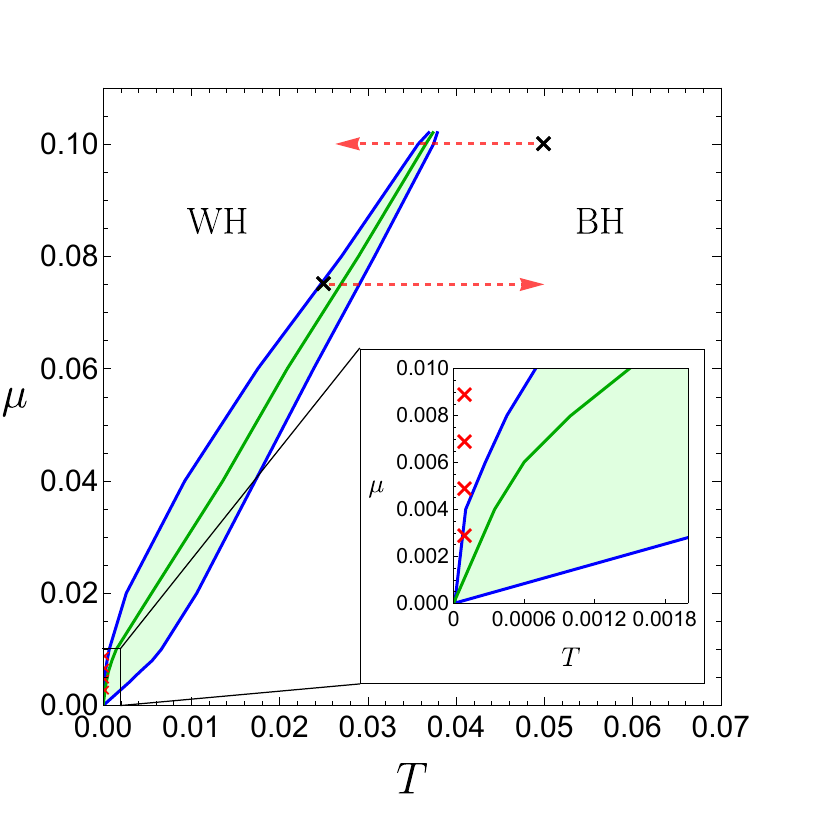}
    \hfill
    \includegraphics[width=0.5\linewidth]{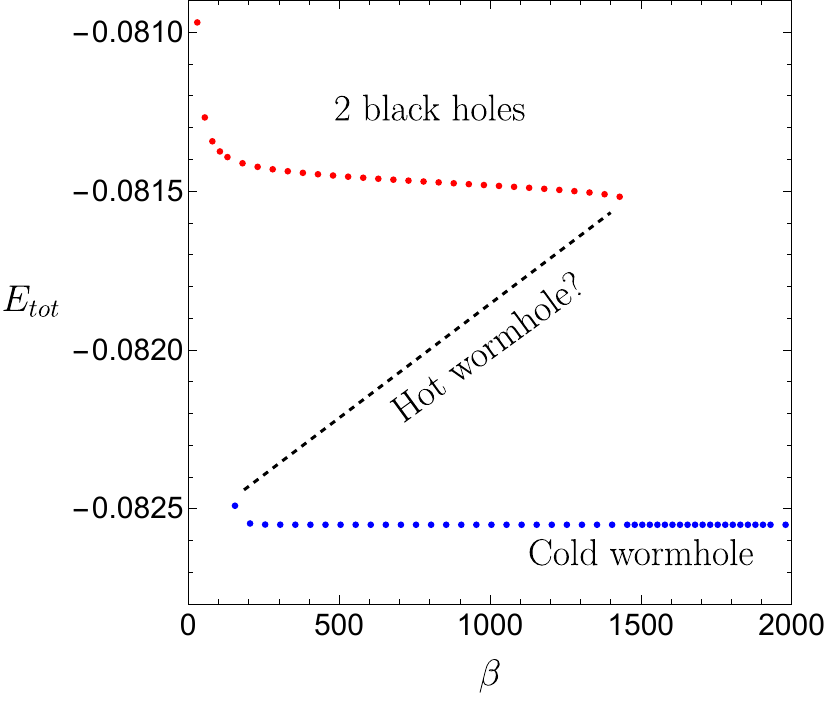}
    \caption[]{Left: $\mu$-$T$ phase diagram separating the two phases. The colored region shows where the two phases coexist. On the left (right) of the green line, the wormhole (black hole) phase is more stable. We show in black crosses the areas of the phase diagram that have been explored in other non-equilibrium analysis of the model \cite{MaldaMilekhin,Zhang_2021}, with arrows showing the trajectories of their processes\footnotemark. We show in red crosses the areas we are able to reach with NESSi, much deeper inside the wormhole regime. Right: phase diagram for $\mu=0.01$. The existence of a canonically unstable but microcanonically stable phase can be seen by computing the energy of each phase.}
    \label{fig:phasediagram}
\end{figure}

Initially, the system is set to a thermal equilibrium state characterized by a temperature parameter, $\beta$, and the drivings are turned on at $t_0=0$. Specifics regarding the derivation of the equilibrium initial conditions can be found in Appendix \ref{App:initialconds}. The predictor-corrector method hinges upon the exponential decay of equilibrium Green's functions, a typical behavior in the black hole phase. However, in the wormhole phase, the revival dynamics significantly slows down the decay, especially at high $\beta$ values which renders it impractical. To circumvent these limitations, leveraging the NESSi (Non-Equilibrium Systems Simulation \cite{Schuler_2020}) library has proven crucial for exploring the most interesting region of the phase diagram $-$specifically, the low-$\mu$, low-$T$ area within the wormhole phase (the lowest-left corner in Fig. \ref{fig:phasediagram}). NESSi adopts the L-shaped Kadanoff Baym contour shown on the right plot in Fig. \ref{fig:Keldyshcontour}, hence obtaining the equilibrium state directly from the Matsubara formalism. Additionally, NESSi incorporates advanced high-order integration routines, vital for achieving high-fidelity results. In order to verify the results shown in this paper, we have used both integration schemes when possible.

\footnotetext{Strictly speaking, the phase diagrams of \cite{Maldacena_2017,Zhang_2021} are not exactly equal as the one showed here because their setups are different: in \cite{Maldacena_2017} they couple the system to a cold bath, and in \cite{Zhang_2021} they consider the complex version of the model. However, the scaling of the numerical complexity with $\mu$ and $T$ is comparable in all of them.}

\subsection{Phase diagram and Schwarzian limit} \label{sec:PhaseDiagram}
The phase diagram of the model shows two phases, separated by a first order phase transition \cite{MaldaQi}. The low temperature phase containing the ground state, which is identified with a traversable wormhole geometry, and the high temperature phase, more similar to two disconnected thermal SYK systems, which is identified with 2 AdS black holes (see Fig. \ref{fig:phasediagram}). In the canonical ensemble both phases are connected by an unstable phase, the so-called hot wormhole. This phase is presumably  stable in the microcanonical ensemble \cite{MaldaMilekhin}.

A key feature of the wormhole phase is the regularly peaked structure of the spectral function, as opposed to the continuum shown in the SYK case. 
As shown in \cite{Plugge_2020}, this spectrum neatly approaches the conformal spectrum $E =\epsilon~(\Delta+n)$ with $\Delta = 1/4$ and $\epsilon \propto \mu^{2/3}$, in the limit $\mu\to 0$. Such regularity is behind the  revival phenomenon that shows up as the dual counterpart to the traversing of the wormhole by a probe particle \cite{Maldacena_2017,Bak:2019nnu}. These revivals are neatly seen in the time evolution of the transmission coefficients
\begin{equation}
    T_{ab}(t)=2\abs{G^>_{ab}(t,0)}~.
\end{equation}

\begin{figure}[ht]
    \centering
    \includegraphics[width=0.49\linewidth]{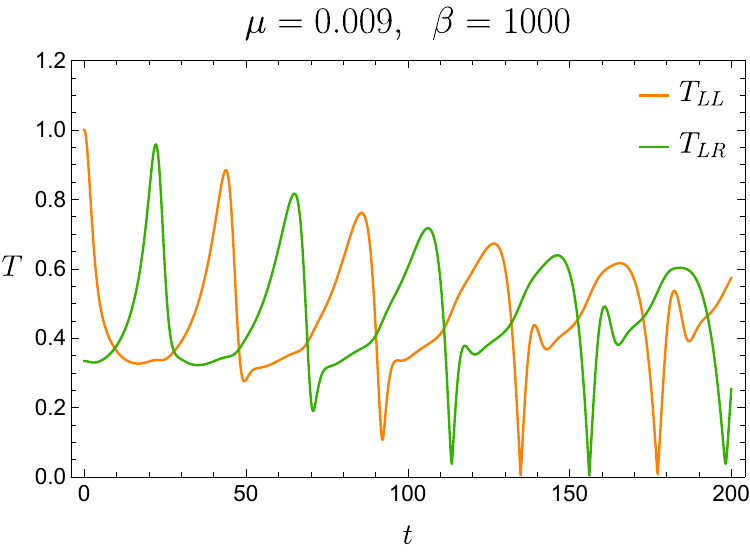}
    \hfill
    \includegraphics[width=0.49\linewidth]{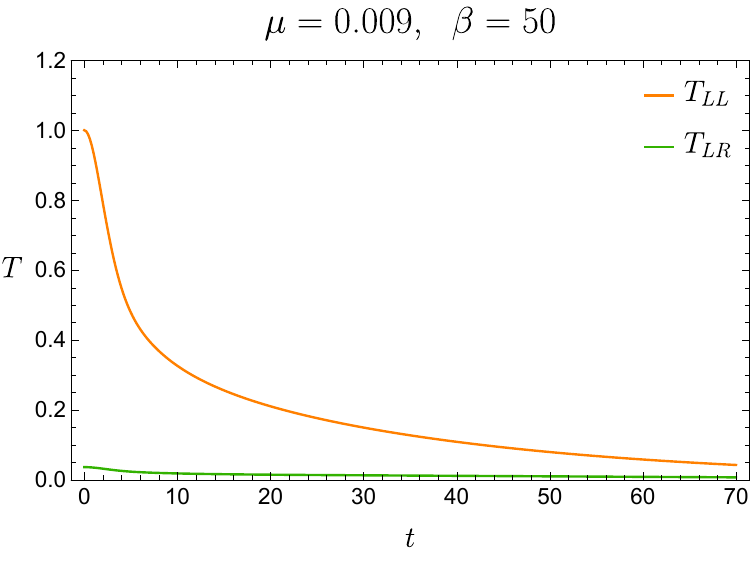}
    \caption{Left (right): transmission coefficients in the wormhole (black hole) phase. The phase opposition of the revivals in the wormhole phase is consistent with having a traversable wormhole. In the black hole phase the correlators decay exponentially with time. Also the frequency scaling as ${\cal O}(\mu^{2/3})$ is, for small $\mu\ll 1$, much higher than the natural coupling frequency ${\cal O}(\mu)$.}
    \label{fig:LHrevivals}
\end{figure}

The transmission coefficients,
$T_{ab}$, encode the probability of recovering $\chi^j_a$ at time $t$  after having inserted $\chi^j_b$ at time $t=0$. Fig. \ref{fig:LHrevivals} shows the different behaviors in the two phases. On the left plot the peaked out of  phase revivals  are interpreted as being duals of a probe particle traversing from one side of the wormhole to the other and bouncing forever back and forth. The  slow decay in the overall envelope still deserves an explanation from the dual gravitational point of view.

Actually, there is another integerly spaced spectrum in the scene: that of the boundary graviton. Let us review how they both come out following closely the reasoning given in \cite{MaldaQi}. 
In the low-energy limit $\mu\ll J$ the model is effectively governed by the Schwarzian action
\begin{equation}
    S=-N\int d\tilde{u}\left\{\left\{\tan\frac{t_l(\tilde{u})}{2},\tilde{u}\right\}+\left\{\tan\frac{t_r(\tilde{u})}{2},\tilde{u}\right\}-\eta\left[\frac{t'_l(\tilde{u})t'_r(\tilde{u})}{\cos^2\frac{t_l(\tilde{u})-t_r(\tilde{u})}{2}}\right]^\Delta \right\}~.
    \label{eq:Schwarzianaction}
\end{equation}
The first two terms correspond to the reparametrization modes arising from the conformal symmetry of each SYK model in the IR limit. The interaction between boundaries introduces the remaining term, with
\begin{equation}
    \eta\equiv \frac{\mu \alpha_S}{\mathcal{J}}\frac{c_\Delta}{(2\alpha_S)^{2\Delta}}\, ,
    \label{eq:etadef}
\end{equation}
with $\mathcal{J}^2=J^2/2$ and $c_\Delta=\frac{1}{2}\left[(1-2\Delta)\frac{\tan \pi\Delta}{\pi\Delta}\right]^\Delta$. In our case, $\Delta=1/4$, $\tilde{u}$ is a rescaled boundary time defined by $\tilde{u}\equiv \frac{\mathcal{J}}{\alpha_S}u$ and $\alpha_S \approx 0.007$ is a numerical constant  \cite{MaldaStanford}.
A simple solution to the equations of motion derived from the action \eqref{eq:Schwarzianaction} is  
\begin{equation}   t_l(\tilde{u})=t_r(\tilde{u})=t'\tilde{u},~~~~\text{with}~~~t'\equiv dt/d\tilde{u}=\left(\Delta\eta\right)^{\frac{1}{2(1-\Delta)}}\, ,
    \label{eq:tpconstant}
\end{equation}
from where two towers of states emerge. One comes from conformal symmetry: a primary operator of dimension $\Delta$ creates one with  energies $E_t=\Delta+n$ with respect to global bulk time. With respect to boundary time $\tilde{u}$ the energies are instead
$E_{\tilde{u}}=t'(\Delta+n)$.

The second tower emerges when one considers all physical solutions also for non-constant $t'(\tilde{u})$. It turns out that one can restrict the attention to solutions of the form $t'_l(\tilde{u})=t'_r(\tilde{u})=t(\tilde{u})$ satisfying the equation (written in terms of $\varphi=\log t'(\tilde{u})$)
\begin{equation}
    -e^{2\varphi}-\varphi''+\Delta\eta e^{2\Delta \varphi}=0~.
    \label{eq:eomvarphi}
\end{equation}
This corresponds to the equations of motion of a non-relativistic particle in a potential with the action
\begin{equation}
    S=N\int d\tilde{u}\left(\Dot{\varphi}^2-V(\varphi)\right),~~~~V(\varphi)=e^{2\varphi}-\eta e^{2\Delta \varphi}~.
\end{equation}

The solution \eqref{eq:tpconstant} corresponds to the minimum of the potential, $e^{2(1-\Delta)\varphi_m}=\Delta\eta$. The oscillations around the minimum give rise to a harmonic oscillator degree of freedom with frequency and physical energies, respectively,
\begin{equation}
    \omega_0=t'\sqrt{2(1-\Delta)},~~~~~~~~~~~~E_{\tilde{u}}=\omega_0 \left(n+\frac{1}{2}\right)~.
    \label{eq:bdrygrav}
\end{equation}
This degree of freedom corresponds to the "boundary graviton".
Finally, physical correlators can be obtained from the knowledge of  $t'$ 
\begin{equation}
    \langle\mathcal{O}(\tilde{u}_1)\mathcal{O}(\tilde{u}_2)\rangle=\left[\frac{t'}{\cos\frac{t'(\tilde{u}_1-\tilde{u}_2)}{2}}\right]^{2\Delta}~.
    \label{eq:propagators}
\end{equation}

\subsection{Observables of interest}

In order to monitor the response of the system to the driving, we will analyze the time evolution of the total energy, $E_{tot}(t)=\langle H\rangle =\langle H_L\rangle+\langle H_R\rangle+\langle H_{int}\rangle$ in the hamiltonian \eqref{eq:tpham} for different frequencies. The energy can be written in terms of the Green's functions, and in the case of arbitrary time-dependent couplings $\mu(t)$ and $J_{L/R}(t)$ it acquires the following form
\begin{equation}
    \frac{1}{N}\langle H_L(t)\rangle=-\frac{iJ_L(t)}{4}\int_{-\infty}^{t}dt'\Big[J_L(t')\left(G_{LL}^>(t,t')^4-G_{LL}^<(t,t')^4\right)+J_R(t')\left(G_{LR}^>(t,t')^4-G_{LR}^<(t,t')^4\right)\Big]~,
    \label{eq:E_L}
\end{equation}
with the corresponding expression for $\langle H_R(t)\rangle$ upon doing $L\leftrightarrow R$. The total energy is given by
\begin{equation}
    E_{tot}(t)=E_L(t)+E_R(t)+E_{int}(t)~,
    \label{eq:Etot}
\end{equation}
where $E_{int}(t)=-N\mu(t)G^>_{LR}(t,t)$. Details are provided in Appendix \ref{App:energies}. 
However, it is evident from the expressions that the time-dependent modulations $J_{L/R}(t)$ as well as $\mu(t)$ will dominate and will "hide" the part of the evolution of the energy that comes from the response of the system to the drivings, encoded in the time evolution of the Green's functions. For that reason, we choose to evaluate these expressions taking the equilibrium values of the couplings $J_{L/R}$ and $\mu$. This corresponds to the expectation values of the undriven hamiltonian, $\langle H\rangle (t)$.\footnote{We thank C. Kuhlenkamp for clarification on this point. }

We would also like to study the heating dynamics. Of course,  out of equilibrium the notion of temperature is ill-defined, but there are some {\it ad hoc}  prescriptions to define an "effective temperature". One standard approach  is based on the observation that, in thermal equilibrium, where all Green's functions depend only on time differences, the following relation in Fourier space holds
\begin{equation}
    \frac{i G^K(\omega)}{\rho(\omega)}=\tanh\frac{\beta\omega}{2}~,
    \label{eq:FDTeq}
\end{equation}
where $G^K(\omega)$ is the Fourier transform of the Keldysh Green's function, defined in \eqref{eq:retadvkel}, and $\rho(\omega)=-2\Im G^R(\omega)$ is the spectral function. This relation follows from the KMS condition
\begin{equation}
    G^>(\omega)=-e^{-\beta\omega}G^<(\omega) \,.
\end{equation}

Equation \eqref{eq:FDTeq} allows to extract the temperature of the system from the Green's functions. 
Out of equilibrium, the Green's functions depend in general on two times, $G(t_1,t_2)$. The standard procedure amounts  to rotating the time variables into a relative time $t$ and an average time $\mathcal{T}$, defined as
\begin{equation}
    t=t_1-t_2,~~~~~\mathcal{T}=\frac{t_1+t_2}{2}~,
\end{equation}
and computing the Fourier transform only with respect to $t$. This yields the Wigner transform $G(\mathcal{T},\omega)$,
\begin{equation}
    G(\mathcal{T},\omega)=\int dt e^{i\omega t}G(t_1=\mathcal{T}+\frac{t}{2},t_2=\mathcal{T}-\frac{t}{2})~.
\end{equation}
Defined  this way, the time-translation symmetry breaking due to the driving is contained in the dependence on $\mathcal{T}$. Now we define the "time-dependent" quantities $G^K(\mathcal{T},\omega)$, $\rho(\mathcal{T},\omega)$, and fit
\begin{equation}
    \frac{i G^K(\mathcal{T},\omega)}{\rho(\mathcal{T},\omega)}=\tanh\frac{\beta(\mathcal{T})\omega}{2}
    \label{eq:betaeff}
\end{equation}
to obtain a time-dependent effective temperature. The caveat is that this quotient will only approach a hyperbolic tangent when the system is not too far from an equilibrium configuration. We will use this $\beta(\mathcal{T})$ as an indicator of the heating. At a numerical level, the Wigner transform cannot be computed in a finite domain of times
unless the decay in relative time $t = t_1-t_2$ is strong enough. In the wormhole phase the oscillations decay at a very slow rate, so this strategy can not be used in general and we will face strong limitations when computing the effective temperature for large $\beta$ (small $\mu$) wormholes. However, the energy is numerically easier to compute in all cases.

\begin{figure}[t]
    \centering
    \includegraphics[width=0.49\textwidth]{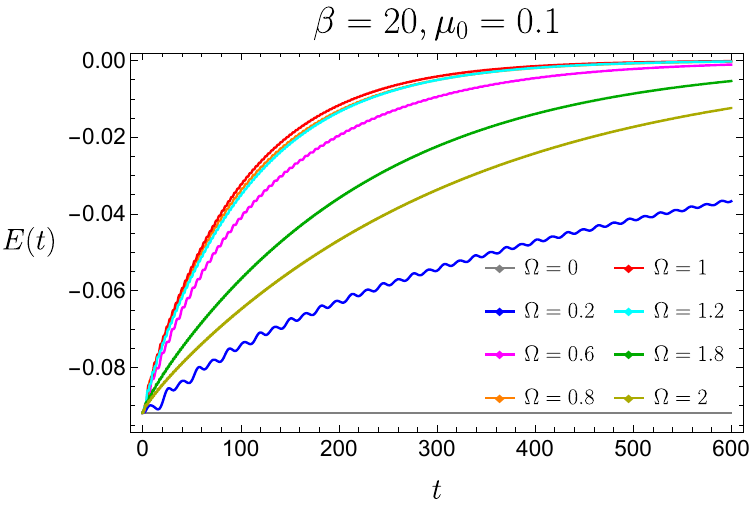}
    \includegraphics[width=0.49\textwidth]{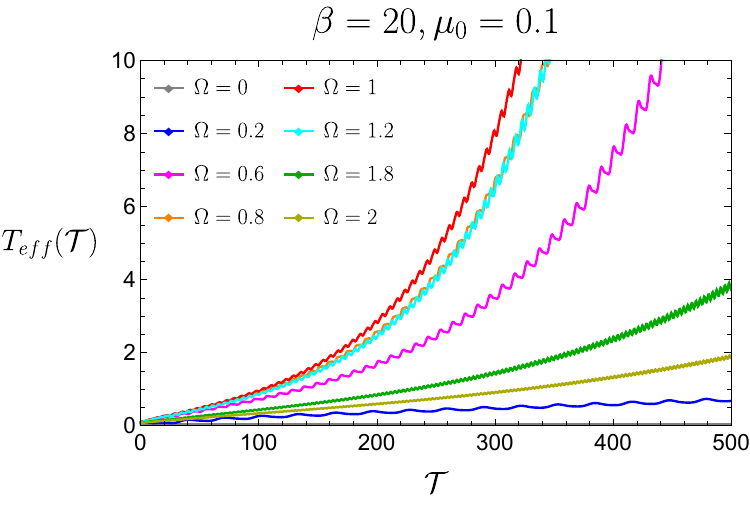}
    \caption{Time dependence of the energy and the effective temperature for different frequencies of the driving in the black hole phase. The amplitude is taken to be $a=0.7$. We observe that the absorption and the heating reach a maximum value for $\Omega\sim 1$ and then they decrease again.}
    \label{fig:BHenergybeta}
\end{figure}

\section{Numerical results}
\label{sec:results}

This section contains the main contributions in this work. When possible we have made two different codes in order to double check the results. In some situations, however, only one type of code was able to provide resources to accomplish the simulation. This is especially true in the wormhole phase. 

\subsection{Driving the \texorpdfstring{$L$}{L}-\texorpdfstring{$R$}{R} coupling \texorpdfstring{$\mu$}{mu}}\label{subsec:mudriving}
We begin our numerical study by perturbing sinusoidally  $\mu$, that is the coupling between the $L$ and $R$ sides appearing in  
\eqref{eq:tpham}
\begin{equation}
        \mu(t)=\mu_0(1+a\sin\Omega t)~.
        \label{eq:driving}
\end{equation}
The most interesting effects arise in the wormhole phase (low $\mu$, large $\beta$), but we begin by considering the black hole phase for later comparison.
Some examples of the $t$-dependent energy, $E_{tot}(t)$,  computed using expressions \eqref{eq:E_L}-\eqref{eq:Etot}, and the ${\mathcal T}$-dependent effective temperature, $T_{eff}({\mathcal T})$, from \eqref{eq:betaeff}, are shown in Fig. \ref{fig:BHenergybeta}.

\begin{figure}[t]
     \centering
     \begin{subfigure}[t]{0.47\textwidth}
         \centering
         \includegraphics[width=\textwidth]{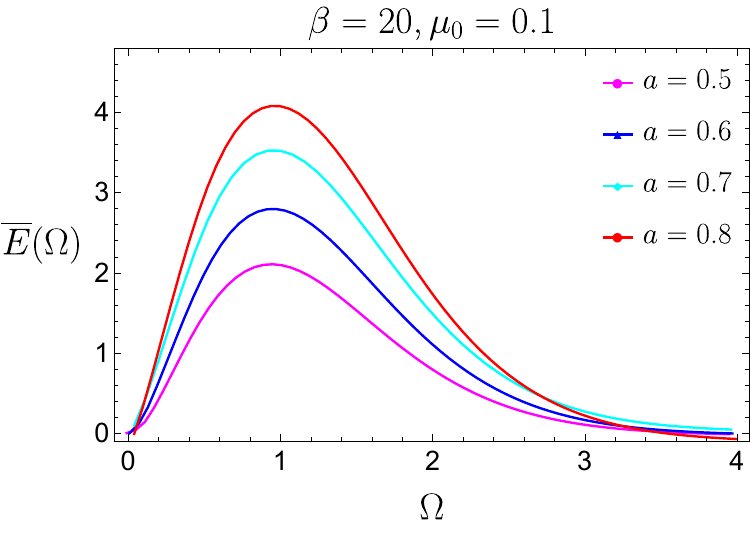}
         \caption{}
         \label{sfig:BHintegr}
     \end{subfigure}
     \begin{subfigure}[t]{0.49\textwidth}
         \centering
         \includegraphics[width=\textwidth]{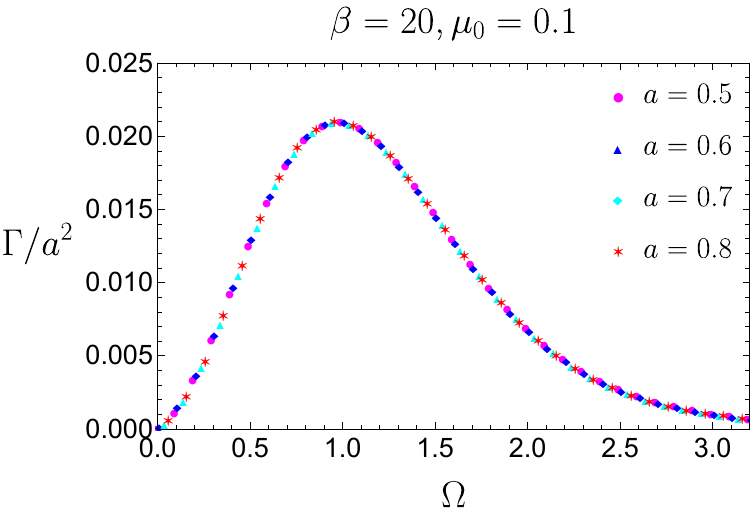}
         \caption{}
         \label{sfig:BHfit}
     \end{subfigure}
        \caption{Left: integrated energy $\overline{E}(\Omega)$ for different amplitudes in the black hole phase. We observe a maximum around $\Omega\sim 1$. Right: heating rate $\Gamma$. We find that the rate has a universal behavior $\Gamma(a,\Omega)=a^2f(\Omega)$, as it was already observed in \cite{Kuhlenkamp_2020}.}
        \label{fig:BHintegrfit}
\end{figure}

In order to characterize and study the net absorption of energy as a function of the driving frequency $\Omega$, we will make use of the integrated energy 
\begin{equation}
    \overline{E}=\int_{0}^{100}\left[E_{tot}(t)-E_{tot}(0)\right]dt~.
\end{equation}
for different driving frequencies. The choice of upper value $t\leq 100$  will become clear in the analysis of the wormhole phase.  Fig. \ref{sfig:BHintegr} shows the value of  $\Bar{E}$ as a function of $\Omega$ for four amplitudes $a$. In all cases the plots  reach maximum around $\Omega\sim 1$.  We fitted the energy to an exponential,
\begin{equation}
    \abs{E_{tot}(t)}\sim e^{-\Gamma(a,\Omega) t}~,
\end{equation}
with a driving-dependent heating rate $\Gamma(a,\Omega)$. Fig. \ref{sfig:BHfit} shows this rate, where we observe a universal behavior, $\Gamma(a,\Omega)=a^2f(\Omega)$, that matches the one  observed in \cite{Kuhlenkamp_2020} in the  strange-metal phase.

New physics appears if the initial state belongs, instead, to the wormhole phase. Consider for instance an initial equilibrium solution with $\beta=1000$, $\mu=0.009$. At $t=0$ we turn on the driving \eqref{eq:driving} and see how the transmission amplitudes between the two sides are affected by the injection of energy. For high amplitudes of the driving ($a\sim \mathcal{O}(1)$) the injection of energy is so violent that the system transitions to the black hole phase almost immediately. To remain in the wormhole phase for longer times it is necessary to choose smaller perturbation amplitudes.

\begin{figure}[htbp]
     \centering
     \begin{subfigure}[t]{0.49\textwidth}
         \centering
         \includegraphics[width=\textwidth]{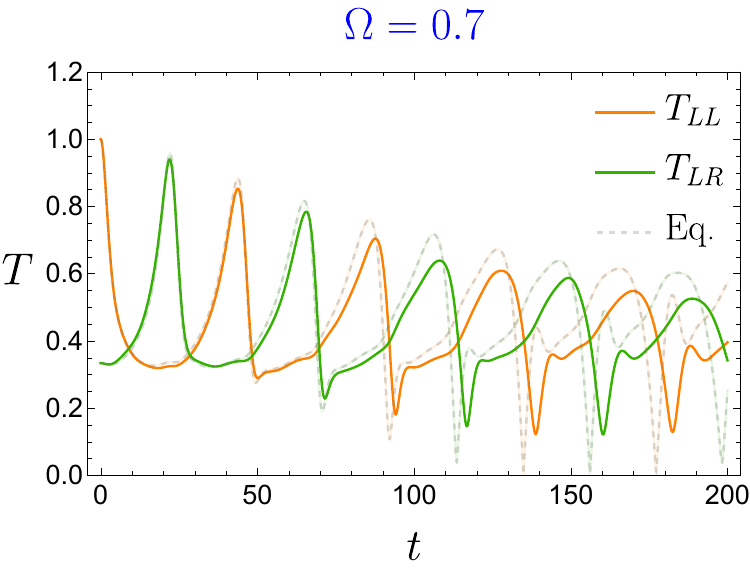}
         \caption{}
         \label{sfig:WHrevivalsnores}
     \end{subfigure}
     \begin{subfigure}[t]{0.49\textwidth}
         \centering
         \includegraphics[width=\textwidth]{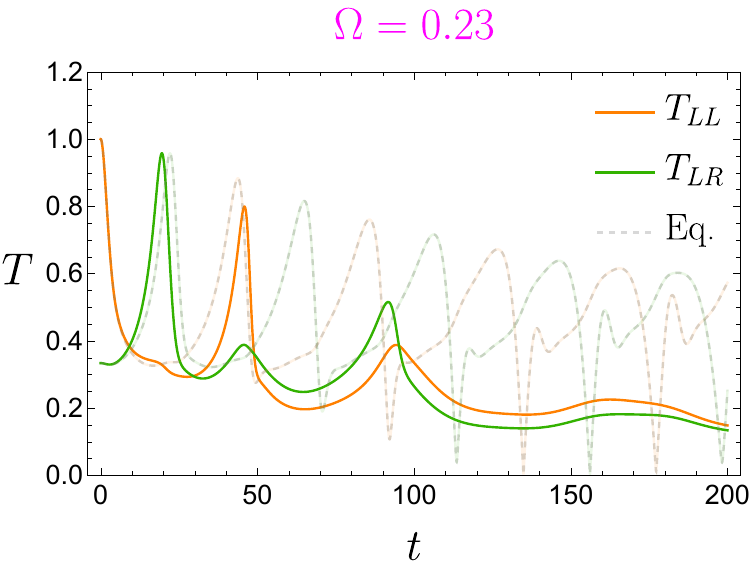}
         \caption{}
         \label{sfig:WHrevivalsres}
     \end{subfigure}
     \begin{subfigure}[t]{0.55\textwidth}
         \centering
         \includegraphics[width=\textwidth]{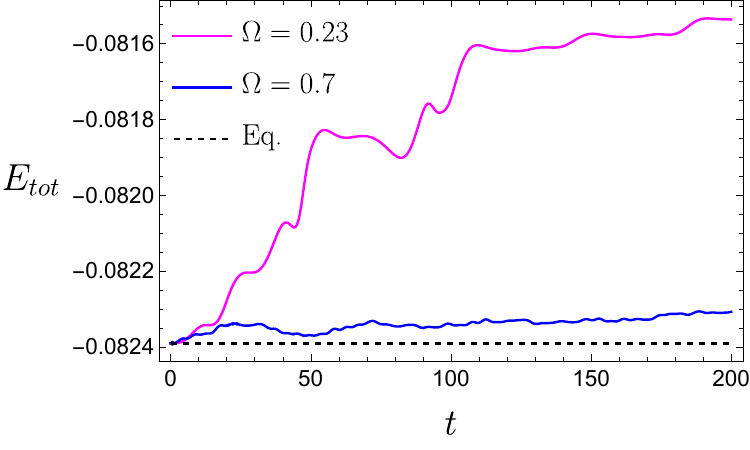}
         \caption{}
         \label{sfig:WHenergy}
     \end{subfigure}
        \caption{The two different behaviors we observe. In \ref{sfig:WHrevivalsnores} for non-resonant frequencies the revivals retain mostly their equilibrium form. In \ref{sfig:WHrevivalsres}, for resonant frequencies we observe a rapid decay of the transmission amplitudes while they cease to be in phase opposition, signalling that the wormhole has closed. For comparison,  we plot the shape of the revivals in equilibrium in light dashed. \ref{sfig:WHenergy}: we show the behavior of the energy for the two cases. We observe how, when driving at the resonant frequencies, the early time absorption becomes exponential. Between $t=50$ and $t=100$, the behavior changes. Looking at \ref{sfig:WHrevivalsres}, this corresponds to the moment where the transition to the black hole phase is happening. After $t=100$, the behavior is essentially the same as in the non-resonant case. }
        \label{fig:WHrevivals}
\end{figure}

When $a$ is small enough, we find two types of behavior upon varying the driving frequency $\Omega$. In most of the frequency range, the revivals are almost unaffected (Fig. \ref{sfig:WHrevivalsnores}). However, for some discrete set of frequencies, the transmission amplitudes decay and become similar to those of the black hole phase (Fig. \ref{sfig:WHrevivalsres}). We interpret this effect as a resonance that triggers a transition to the high-temperature phase.

One would like to understand the nature of the resonant frequencies. As we show in Fig. \ref{sfig:WHenergy}, the absorption of energy is very different between resonant and non-resonant frequencies: while in the non-resonant case the energy increases very slowly, in the resonant frequencies the heating is exponential at early times, triggering the transition to the high-temperature (black hole) phase. In section \ref{sec:hot_wormhole} we analyze this transition in more detail.

\begin{figure}[!ht]
     \centering
     \begin{subfigure}[t]{0.49\textwidth}
         \centering
         \includegraphics[width=\textwidth]{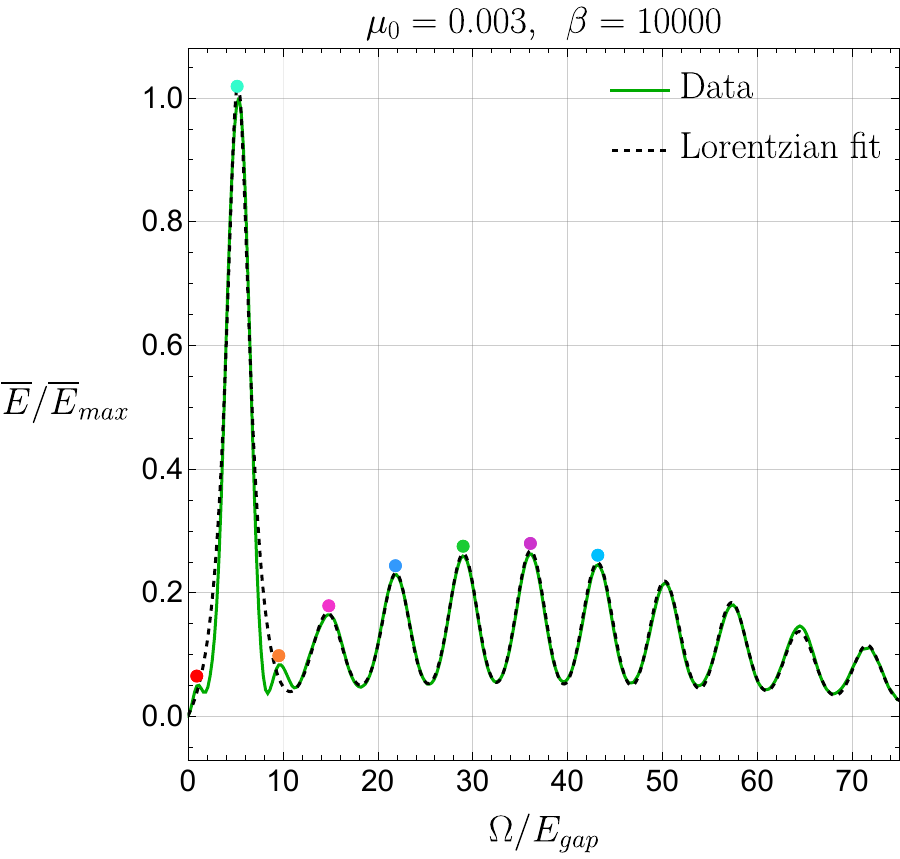}
         \caption{}
         \label{sfig:WHintegrenergy}
     \end{subfigure}
     \begin{subfigure}[t]{0.49\textwidth}
         \centering
         \includegraphics[width=\textwidth]{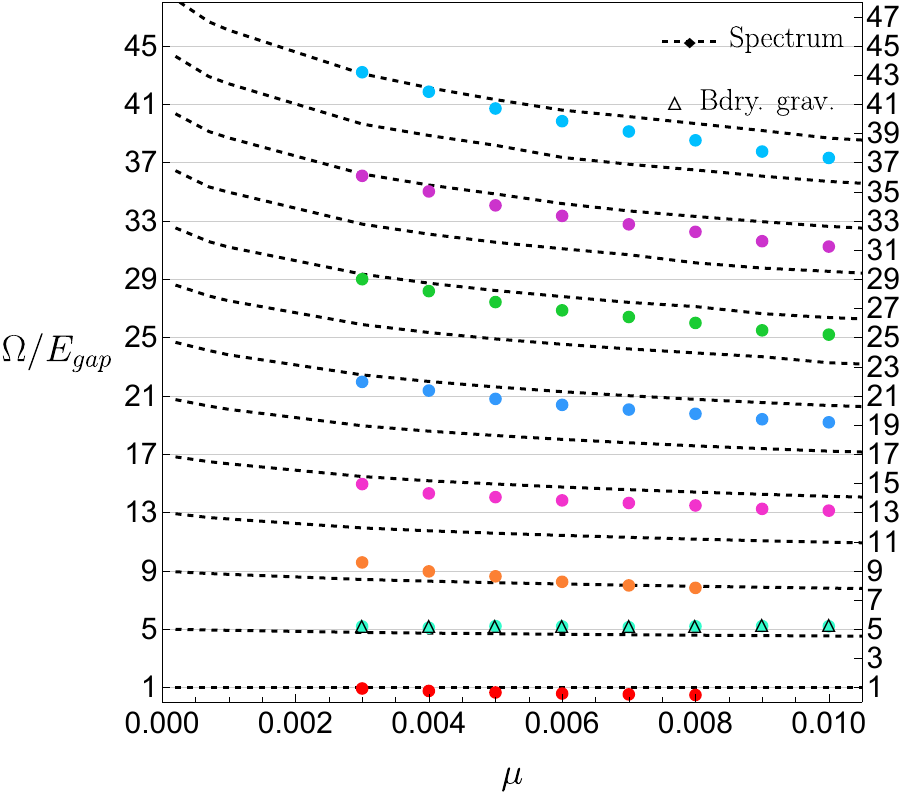}
         \caption{}
         \label{sfig:Peaksspectrum}
     \end{subfigure}
        \caption{Left:~(normalized) integrated energy for an amplitude $a=1/50$. We observe a series of absorption peaks at the frequencies where the revivals are suppressed. The peaks are well approximated by a sum of Lorentzians. Right: Solid dots mark absorption peaks for varying $\mu$,  with the same color coding as in the left plot. In dashed, a fit to the  spectrum of the undriven model computed numerically in \cite{Plugge_2020}. We also show the frequency $\omega_0(\mu)$ of the boundary graviton (see equation \eqref{eq:bdrygrav}), which coincides with our second (and highest) absorption peak on the left.}
        \label{fi}
\end{figure}

The dependence of the  energy absorption,  $\bar E$, with the driving frequency, $\Omega$, departs significantly from the one in the black hole phase shown in Fig. \ref{sfig:BHintegr}. Now, Fig. \ref{sfig:WHintegrenergy} exhibits the integrated energy as one varies the frequency\footnote{For this, we choose an amplitude small enough to remain in the wormhole phase during all the time used for the integrated energy ($t=100$). In this way we capture the intrinsic behavior of the wormhole phase, and we don't get contamination from the transition to the black hole phase, which will be reached inevitably at some later time.}.    The peaks correspond precisely to the frequencies where the revivals get maximally suppressed and the overall curve is well fitted to a sum of Lorentzians
\begin{equation}
    \overline{E}(\Omega)\sim \sum_{i=\text{peaks}}\frac{1}{\pi}\frac{\gamma_i}{(\Omega-\Omega_i)^2+\gamma_i^2}~,
\end{equation}
with $\gamma_i$, $\Omega_i$ being characteristic of each resonant peak.\footnote{The global fit works better if we don't include the two small peaks before and after the highest one. However, we consider them to be resonances.}

The resonant frequencies are easily extracted from Fig. \ref{sfig:WHintegrenergy}, but the question of their meaning remains open. For that, we repeated the analysis for lower values of $\mu$. In each case, we compute the integrated energy and identify the resonant frequencies as the frequencies where the absorption has a maximum and the corresponding transmission coefficients are highly suppressed. After all the resonances are collected, the natural thing to do is to compare them with the spectrum of the undriven system, which was computed numerically in \cite{Plugge_2020}.

From results,  shown in Fig. \ref{sfig:Peaksspectrum}, one observes that the resonances indeed come very close to the levels of the spectrum. Remarkably, this only happens in about half of the spectrum. This is a very intriguing feature for which we don't have a clear explanation. Presumably, this selection rule should come from the structure of the excitations, most likely the fermion number of ${\cal O}(N)$ invariant states. To study this phenomenon from the dual gravity side, a fully backreacted gravitational calculation in the dual geometry seems to be needed. Following the calculations in \cite{Gao_2017, Bak:2018txn}, one could study how the violation of the averaged null energy condition is affected by the driving.

The second resonance stands out prominently. Although it is close to the second level ($n=1$) of the conformal spectrum, it would be the only peak that violates the evenly spaced structure found for the others. 
In fact, it seems to be closer to the {\it natural frequency} of the boundary graviton, $\omega_0$ given in \eqref{eq:bdrygrav}. This is surprising at first sight, since graviton states don't seem to show up as features of the spectral function \cite{Plugge_2020}. This fact was justified from being expected only as a $1/N$ order effect beyond the saddle point calculation involved here. 
In contrast, the fact that we can see it could come from driving, precisely, the parameter that  gives rise to this spectrum. We will give more support to this claim below in  Section \ref{sec:schwarz},  analytically in the limit where the Schwarzian approximation is reliable.

\subsection{Forming hot wormholes?}
\label{sec:hot_wormhole}

It has been  argued  that the black hole  and  wormhole phases are connected by a set of metastable states referred to as "hot wormholes" \cite{MaldaQi,MaldaMilekhin}. Being unstable  in the canonical ensemble, one does not expect to find them  when solving the equilibrium Schwinger Dyson equations at a given temperature. By putting the system in thermal contact with a cooler bath, evidence of the presence of those states is obtained from the behaviour of the temperature with energy \cite{MaldaMilekhin}.

In Fig. \ref{fig:WHrevivals}, we showed that by driving periodically $\mu$ in a resonant frequency it is possible to make the wormhole transition into a black hole. In this section we look for a better characterization of that transition, in the seek for signals that might correspond to such hot wormhole states.

\begin{figure}[h!]
     \centering
     \begin{subfigure}[t]{0.55\textwidth}
         \centering
         \includegraphics[width=\textwidth]{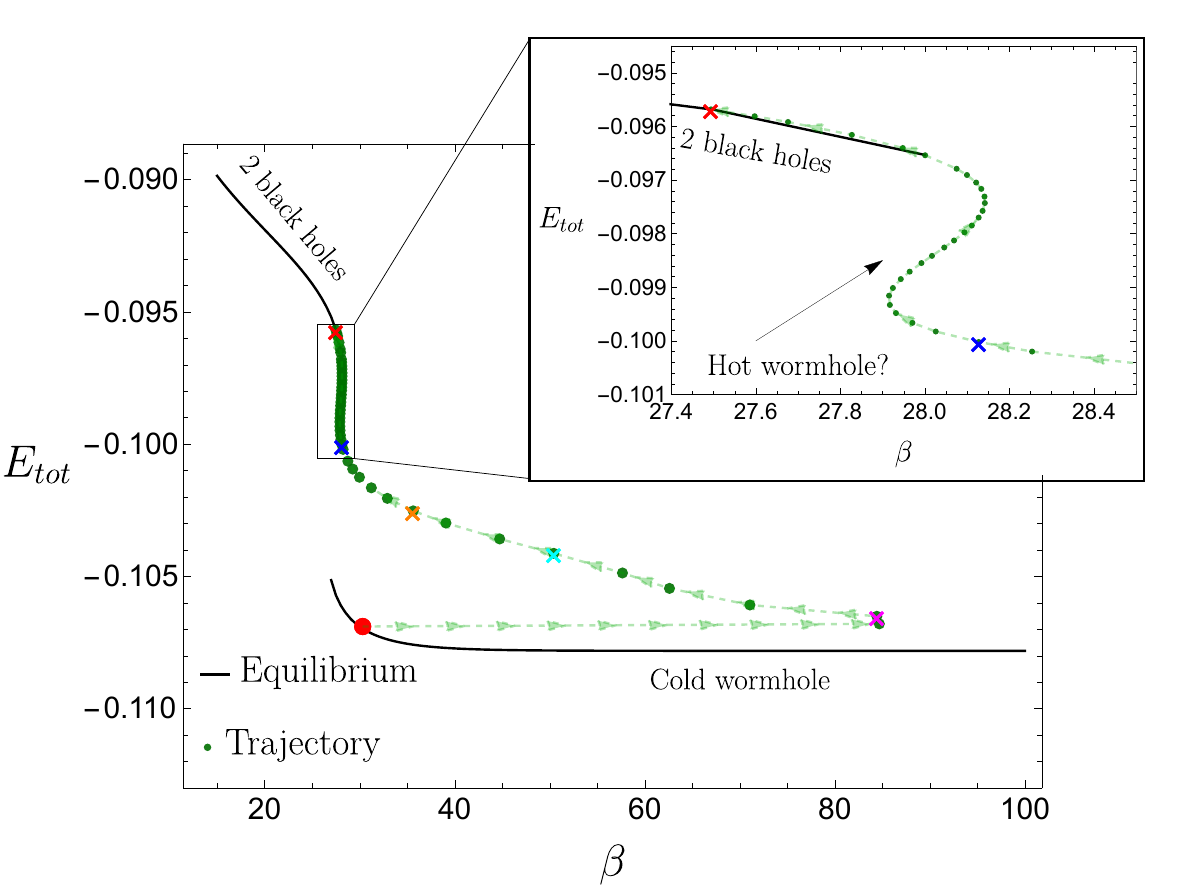}
         \caption{}
         \label{sfig:hotWHdiag}
     \end{subfigure}
     \begin{subfigure}[t]{0.44\textwidth}
         \centering
         \includegraphics[width=\textwidth]{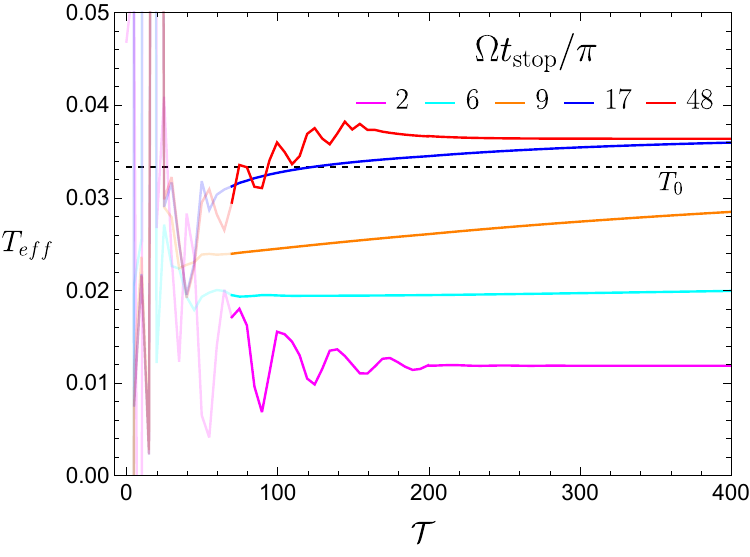}
         \caption{}
         \label{sfig:hotWHTeff}
     \end{subfigure}
        \caption{Left: asymptotic equilibrium trajectory in the phase diagram. The initial equilibrium solution (red dot) is a wormhole solution of $\mu=0.1$, $\beta=30$, very close to the transition point. We take the drivings to have amplitude $a=0.1$ and frequency $\Omega=1$, and each green dot represents the final equilibrium temperature and energy for growing $t^n_{stop}=\frac{\pi}{\Omega}n$ along the direction of the green arrows. We see that the system follows a trajectory that joins smoothly with the black hole phase. Zooming at the points with larger $t_{stop}$, we see that the trajectory has a shape that is characteristic of a phase transition between the phases, as expected. Right: time evolution of the effective temperature for the highlighted points with colored crosses of Fig. \ref{sfig:hotWHdiag}.}
        \label{fig:hotWH}
\end{figure}

Our protocol starts by initializing a wormhole state at equilibrium near threshold. Then we activate the driving as in \eqref{eq:driving}, and sustain it for a certain number of half cycles, $t_{stop}=\frac{\pi}{\Omega}n$. After this, the driving is quenched without discontinuities, and the system returns smoothly to isolation. As a consequence of the driving, the energy of the system will have risen in a way proportional to $n$, and remain as a constant of motion after switch-off. Effectively we are in the microcanonical ensemble and we may continue the free evolution to look for late time thermalization. After equilibrium has settled, the effective temperature of the state can be determined using the aforementioned method \eqref{eq:betaeff}. Each end state gives a point in an $E$-$\beta$ plot.  By repeating this procedure with increasing values of $t_{stop}$, the obtained points trace out the green curve on the left plot in Fig. \ref{sfig:hotWHdiag}.

Monitoring the evolution of the effective temperature down to equilibration requires that the Green's functions decay rapidly enough at the boundaries of the time domain.
 This is numerically demanding in the wormhole phase where the Green's functions decay slowly for small $\mu$ and large $\beta$.  This puts restrictions in these parameters in order to make  the numerics fit within our resources: we will chose for our initial equilibrium state a wormhole with $\beta=30$, $\mu=0.1$. This is close to the phase transition temperature $\beta\approx27.5$ but far from the small $\mu$ large $\beta$ limit considered in previous sections.

In Fig. \ref{fig:hotWH} we present the results of this driving protocol. We show in black the equilibrium phase diagram (see Fig. \ref{fig:phasediagram} right). Each green dot represents a different simulation, with a different value of $t_{stop}$, all starting in the same initial configuration in the wormhole phase represented by the red dot. The dashed line with arrows shows the direction of increasing $t_{stop}$. We highlight five points in the trajectory with color crosses for which we show, on the right plot, the evolution of the effective temperature as computed using \eqref{eq:betaeff}. We observe that for small values of $t_{stop}$, the system  decreases notably its final  temperature. For larger values of $t_{stop}$, the system follows a trajectory that joins smoothly with the equilibrium curve of the black hole phase. 

In summary, whereas driving the system periodically triggers a  transition into the black hole phase (see previous section), a quenched version of the protocol followed by later free evolution brings the system into some stationary state that we propose to correspond to the conjectured hot-wormhole state. 
The states with high $\beta$ obtained for small values of $t_{stop}$ (magenta, light blue and orange points) are equilibrium states that deserve some further study which  will be undertaken in a forthcoming work. 

\subsection{Driving the source couplings \texorpdfstring{$J_{L,R}(t)$}{J(t)} }

Next we shall report on our findings for a periodic driving of the other couplings in the time dependent Hamiltonian  \eqref{eq:tpham}, namely the source couplings $J_{L,R}(t) =J f_{L,R}(t)$. This driving is less motivated physically from an operational point of view as it implies a periodic variation of the variance of the Gaussian distribution of couplings in each SYK.  
Our analysis will be more qualitative than the one carried out for the driving in $\mu(t)$. A full and thorough study will be reported elsewhere.

\begin{figure}[h!]
     \centering
     \begin{subfigure}[t]{0.49\textwidth}
         \centering
         \includegraphics[width=\textwidth]{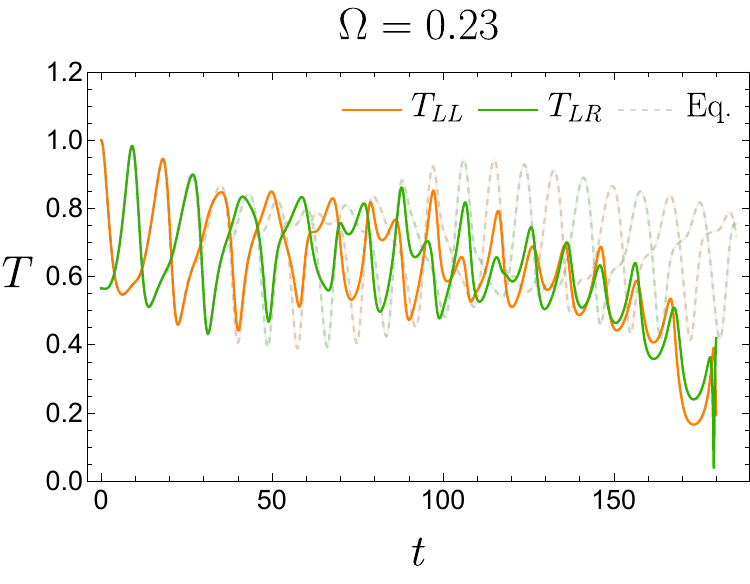}
         \caption{}
         \label{sfig:WHJLdepl}
     \end{subfigure}
     \begin{subfigure}[t]{0.49\textwidth}
         \centering
         \includegraphics[width=\textwidth]{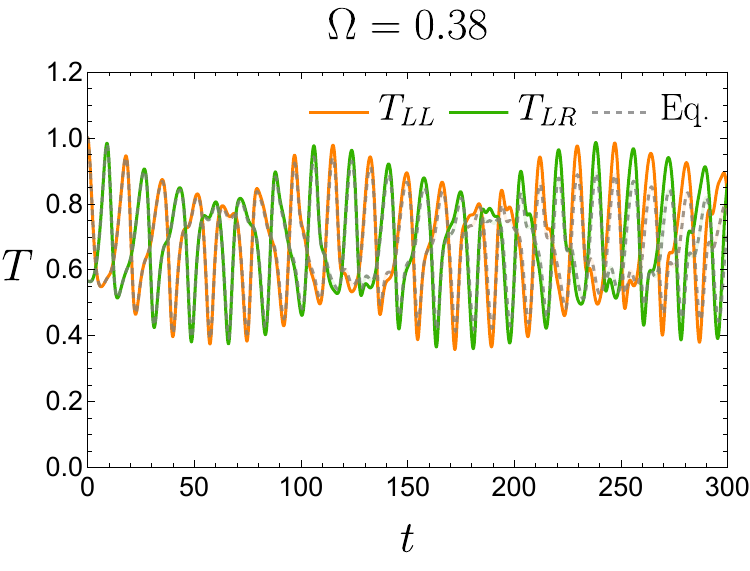}
         \caption{}
         \label{sfig:WHJLJRenh}
     \end{subfigure}
        \caption{Left: Depletion of the transmission coefficients under the asymmetric driving $J_L(t) = J+a\sin \omega t$ while $J_R$ stays constant.  Right: enhancement of the transmission under the driving $J_L(t) = J+a\sin \omega t$ and $J_R(t) = J-a\sin \omega t$. In both cases, the initial equilibrium solution has $\mu=0.05$, $\beta=100$.}
        \label{fig:WHdriveJL}
\end{figure}

We solved the Kadanoff-Baym equations for two different cases. In both of them $J_L(t) = J  + a\sin \omega t$ for the $L$ sector after $t=0$. For the $R$ sector, in one case we will leave  $J_R$ constant. In the other we will impose   $J_R(t) = J  - a\sin \omega t$, thus a time periodic generalization of the imbalanced interactions case studied in \cite{Haenel_2021}. The main features we have encountered are essentially two, both of them showing up in both types of driving. 

On one hand, at late times the evolution is always unstable. The total energy grows unboundedly and the numerics break down before we can reach the  endpoint of this instability. This instability gets enhanced by either increasing the amplitude of the perturbation, or by driving at certain resonant frequencies. In Fig. \ref{sfig:WHJLdepl} we see a  behaviour similar to the one found for the driving in $\mu$ for a sinusoidal perturbation of just $J_L(t)$ at a very particular value of its frequency,  leaving $J_R$ constant. As mentioned, the difference here is that also numerical convergence is lost and simulating cannot be continued to late times. This was essential in previous sections to establish the nature of the effect as a transition to the black hole phase. 
\begin{figure}[h!]
    \centering
    \includegraphics[width=0.6\textwidth]{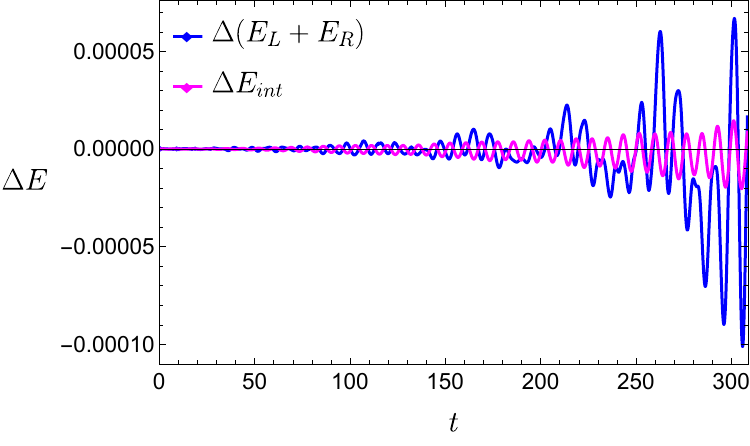}
    \caption{Here we plot the gain in energy, reflecting that we are close to a resonant frequency. The gain occurs in all the components of the total energy, including the interaction part. This is a physical instability that drives the system to another phase which we cannot track up to the end in our simulations because the numerical convergence is lost. Hence we only show the simulation in the reliable interval. }
    \label{fig:WHJLen}
\end{figure}

For  sufficiently small perturbations $a\ll 1$, we have found an intriguing  effect at early times, much before the simulations stop being reliable: an enhancement in the transmission. To get a grasp on this effect it is necessary to recall that, in the undriven case, there is a monotonic drop in the maximum of the oscillations of the transmission coefficients that eventually would lead to a thermalization of the excitation a  late times, exponential in the inverse temperature $\beta$ \cite{Plugge_2020}. The nature of this dissipative mechanism is obscure from the dual gravity point of view and deserves a study.   In Fig. \ref{sfig:WHJLJRenh} we have compared the free evolution with that of a simultaneous driving of $J_L$ and $J_R$ with $a/J = 0.0015$, hence, extremely tiny. For times where the simulation is still reliable way before the instability sets in, $t<300$, a transient period can be seen where an enhancement of the transmission coefficients $T_{LR}$ occurs.

\section{Schwarzian analysis}\label{sec:tdepschwarzian}
\label{sec:schwarz}

We want to study the Schwarzian limit under our periodic driving with the goal of understanding the nature of the dominant peak observed in Section \ref{subsec:mudriving}. In order to do this, we need to make the  replacement $\eta~\rightarrow~\eta(\tilde{u})$ in the action \eqref{eq:Schwarzianaction}, which translates into the same replacement in the equation of motion \eqref{eq:eomvarphi}:
\begin{equation}
    -e^{2\varphi}-\varphi''+\Delta\eta(\tilde{u}) e^{2\Delta \varphi}=0~.
    \label{eq:schweqApp}
\end{equation}

The general solution has to be found numerically, but we can predict the existence of at least one resonance in the limit that the driving is a small perturbation. We consider a driving of the form
\begin{equation}
    \eta~\rightarrow~\eta(\tilde{u})=\eta_0(1+a f(\tilde{u}))~,
\end{equation}
where $\eta_0$ is the equilibrium value, \eqref{eq:etadef}, and we take $a f(\tilde{u})\ll 1$ (at the end we will consider $f(\tilde{u})=\sin\Omega \tilde{u}$, so we will only need $a\ll1$). If we now assume that the deviation from the equilibrium value $\varphi_0$ will also be small, $\varphi(\tilde{u})\approx \varphi_0+\phi(\tilde{u})$, with $\abs{\phi(\tilde{u})}\ll 1$, at first order in $a f(\tilde{u})$ and $\phi(\tilde{u})$, equation \eqref{eq:schweqApp} becomes that of a forced harmonic oscillator
\begin{equation}
    \phi''(\tilde{u})+\omega_0^2\phi(\tilde{u})=\frac{\omega_0^2}{2(1-\Delta)}af(\tilde{u})~,
    \label{eq:forcedoscillator}
\end{equation}
with $\omega_0$ the frequency of the oscillations that give rise to the boundary graviton excitations, \eqref{eq:bdrygrav}, and the driving of $\eta$, $a f(\tilde{u})$, acting as the external force. The solution of \eqref{eq:forcedoscillator} for  general $f(\tilde{u})$ is well known. In the particular case of a sinusoidal driving, $f(\tilde{u})=\sin\Omega \tilde{u}$, it takes the form
\begin{equation}
    \phi(\tilde{u})=\frac{a~\omega_0}{2(1-\Delta)}\frac{1}{\Omega^2-\omega_0^2}\Big[\Omega\sin\omega_0\tilde{u}-\omega_0\sin\Omega\tilde{u}\Big]~,
\end{equation}
which shows a resonance when $\Omega\rightarrow\omega_0$.\footnote{The limit $\Omega\rightarrow\omega_0$ leads to the usual oscillation with a lineraly growing term, $$\lim_{\Omega\rightarrow\omega_0}\phi(\tilde{u})=\frac{a}{4(1-\Delta)}\Big[\sin\omega_0\tilde{u}-\omega_0\tilde{u}\cos\omega_0\tilde{u}\Big]$$} Of course, in that limit the assumption that $\abs{\phi(\tilde{u})}\ll 1$ breaks down and the solution is not valid anymore, but the existence of a resonance in $\phi(\tilde{u})$ (and therefore in $\varphi(\tilde{u})$) will translate into a noticeable change of behavior in the propagators (see \eqref{eq:propagators} and recall that $\varphi(\tilde{u})=\log t'(\tilde{u})$). This analysis confirms that the highest resonant peak we observe corresponds to an excitation of the boundary graviton degrees of freedom.

\section{Summary and conclusions}

In this paper we studied the out of equilibrium dynamics of a system of two coupled SYK models which is  holographically dual to a traversable wormhole in AdS$_2$. In our setup we consider a time-dependent coupling between sides, $\mu(t)$, as well as time-dependent interaction strengths on each of the SYK's, $J_{L,R}(t)$. Our main tool are the Kadanoff-Baym equations (\ref{eq:KBeqs}) for the non-equilibrium Green's functions, which we integrate numerically.

Under the action of the drivings the system absorbs energy and gets heated exponentially. In the case of the driving $\mu(t)$, our numerical results show a clear enhancement of the heating and a depletion of the transmission coefficients for some discrete set of resonant frequencies of the driving. We identified these resonant frequencies with half of the conformal tower of excitations. We also find a pronounced peak for which we provide numerical and analytical evidence as being a resonance coupling the driving to the boundary graviton of the dual theory. We also explored the possibility of having a non-equilibrium phase, the hot wormhole \cite{MaldaMilekhin}, connecting the black hole and the wormhole. Our  results in section \ref{sec:hot_wormhole} provide numerical evidence in favor of the existence of this intermediate phase. 

We have also integrated the non-equilibrium equations for drivings involving the couplings $J_L(t)$ and $J_R(t)$.  The late time simulations become unstable both physically and numerically. For small enough amplitudes of the perturbations, early time evolution indicates a transient enhancement of the signal. 

There are many possible new directions worth exploring in the near future. The asymmetric driving $J_L, J_R$ investigated in this work is not particularly appealing from the physical point of view, as it involves manipulating the variance of the gaussian distribution of couplings. A different option would amount to  inserting a modulated operator on one side and monitoring the effectit causes on the other. One can, for example, add a periodic mass deformation as the one introduced in \cite{kourkoulou2017pure} on one of the sides of the wormhole.   

Another interesting option would be to stabilize heating of the system in order to keep the wormhole open as much time as possible. One possible way to achieve this would be to couple the system to a cold bath that could dissipate the excess heat generated by the driving. The expectation would be to reach in this way a non-equilibrium steady state. A related problem is the study of the transport properties of driven SYK islands connected to thermal reservoirs \cite{Kruchkov_2020,larzul2022energy}. Also to understand the connection of our periodic driving in $\mu$ to the repeated measurement mechanism proposed in \cite{milekhin2023} would be very interesting.

A detailed gravity description of the time-dependent processes studied here is clearly desirable. At low energy we could employ the Schwarzian theory with a periodic time-dependent coupling, completing the analysis of section \ref{sec:schwarz}. Moreover, our setup should also be described in a AdS$_2$ JT gravity theory with a suitable bulk field. In particular, it would be interesting to determine how the driving changes the violation of the average null energy condition with respect to values calculated in \cite{Bak:2018txn, Gao_2017}.

It is also interesting to analyze the entanglement entropy between the two SYK systems when their coupling is periodically driven. One could use for this study the Schwarzian model as in 
\cite{Chen_2019}. A natural question is whether or not there are entanglement resonances occurring at particular frequencies, as those observed in \cite{Sauer_2012} for multipartite quantum systems. 

We expect to have results to report on these issues in the near future.

\section*{Acknowledgements}

We would like to thank J.J. Blanco-Pillado, Elena C\'aceres,  Lata K. Joshi, Clemens Kuhlenkamp, \'Etienne Lantagne-Hurtubise,  Ancel Larzul, Juan Maldacena, Stephan Plugge, Lucas S\'a, Marco Schir\'o, Michael Sch\"uler and  Gustavo Turiaci,  for stimulating and helpful discussions.

Simulations in this work were made possible through the access granted by the Galician Supercomputing Center (CESGA) to its supercomputing infrastructure. The supercomputer FinisTerrae III and its permanent data storage system have been funded by the Spanish Ministry of Science and Innovation, the Galician Government and the European Regional Development Fund (ERDF).
This work has received financial support from Xunta de Galicia (Centro singular de investigaci\'on de Galicia accreditation 2019-2022, and grant ED431C-2021/14), by European Union ERDF, and by the ``Mar\'\i a de Maeztu" Units of Excellence program MDM-2016-0692 and the Spanish Research State Agency (grant PID2020-114157GB-100).

The work of MB has been funded by Xunta de Galicia through the Programa de axudas \'a etapa predoutoral da Xunta de Galicia (Conseller\'ia de Cultura, Educaci\'on e Universidade) and the grant 2023-PG083 with reference code ED431F 2023/19. The work of J.SS. was supported by MICIN through the European
Union NextGenerationEU recovery plan (PRTR-C17.I1), by the Galician Regional Government through the
``Planes Complementarios de I+D+I con las Comunidades Aut\'onomas" in Quantum Communication, and by MCIN/AEI/10.13039/501100011033 and FSE+ with the grant PRE2022-102163.

\appendix

\section{Effective action and Kadanoff-Baym equations}\label{App:KBequations}
We consider the following time-dependent version of the hamiltonian proposed in \cite{MaldaQi},
\begin{equation}   
    H(t)=\sum_{a=L,R}\frac{1}{4!}\sum_{ijkl}f_a(t)J_{ijkl}\chi_a^i\chi_a^j\chi_a^k\chi_a^l+i\mu(t)\sum_j \chi_L^j\chi_R^j~,
\end{equation}
 which couples two identical SYK models with a bilinear term and relevant coupling $\mu$, which now we take to be time-dependent: $\mu(t)$. We also allow for a time dependence in the interaction term of each SYK. $\chi_a^i$ are Majorana operators satisfying the usual anticommutation relations $\left\{\chi_a^i,\chi_b^j\right\}=\delta^{ij}\delta_{ab}$ and $J_{ijkl}$ are real constants drawn from a gaussian distribution with mean and variance given by
\begin{equation}
    \overline{J_{ijkl}}=0,~~~~~\overline{J_{ijkl}^2}=\frac{3! J^2}{N^3}~.
\label{eq:Jmeanvarapp}
\end{equation}

For a general function $f(J_{ijkl})$, the disorder average over the couplings is defined as
\begin{equation}
    \overline{f}(J_{ijkl})=\int\mathcal{D}J_{ijkl} f(J_{ijkl})~,
\end{equation}
where in the case of a gaussian distribution, $\mathcal{D}J_{ijkl}$ is
\begin{equation}
    \mathcal{D}J_{ijkl}=\left(\prod_{ijkl}\frac{d J_{ijkl}}{\sqrt{2\pi\alpha}}\right)\exp\left[-\frac{1}{2\alpha}\frac{1}{4!}\sum_{ijkl}J_{ijkl}^2\right]
\label{eq:DJijkl}
\end{equation}
with $\alpha=\frac{3! J^2}{N^3}$. The averaged partition function is written as\footnote{Here we adopt the annealed average, where one averages first the partition function and computes the expectation values from $\overline{Z}$. This is to be contrasted with the quenched average, where one computes the partition function with the replica method, and then averages the free energy $\overline{\log Z}$. The latter is prefered in general, but in the large-$N$ limit the differences between both methods only gives $1/N$ corrections \cite{Sachdev_2015}.}
\begin{equation}
    \overline{Z}=\int\mathcal{D}\chi_L\mathcal{D}\chi_R\mathcal{D}J_{ijkl}~e^{i S[\chi_L,\chi_R]}~,
\end{equation}
with the action
\begin{equation}
    S=\int_\mathcal{C}dt\Biggl[\frac{i}{2}\sum_{a=L,R}\sum_j\chi_a^j\partial_t\chi_a^j-\frac{1}{4!}\sum_{a=L,R}\sum_{i,j,k,l}f_a(t) J_{ijkl}~\chi_a^i\chi_a^j\chi_a^k\chi_a^l-\frac{i\mu(t)}{2}\sum_j\left(\chi_L^j\chi_R^j-\chi_R^j\chi_L^j\right)\Biggr]
\end{equation}
integrated along the Keldysh contour $\mathcal{C}$ (Fig.\ref{fig:Keldyshcontour}). The gaussian integral over the couplings can be directly evaluated to be

\begin{align}
\begin{split}
    \int \mathcal{D}J_{ijkl} \exp\Biggl[-\frac{1}{4!}\sum_{a=L,R}\sum_{i,j,k,l}\int_{\mathcal{C}}&  dt f_a(t)J_{ijkl} ~\chi_a^i\chi_a^j\chi_a^k\chi_a^l\Biggr]\\
    &=\exp\left[-\frac{N}{2}\frac{J^4}{4}\sum_{a,b}\int_{\mathcal{C}} dt_1~dt_2 f_a(t_1)f_b(t_2)\left[O_{ab}(t_1,t_2)\right]^4\right]~,
\end{split}
\end{align}
where we have defined the bilocal field $O_{ab}(t_1,t_2)$ as
\begin{equation}
    O_{ab}(t_1,t_2)=-\frac{i}{N}\sum_j\chi_a^j(t_1)\chi_b^j(t_2)~.
\end{equation}
In terms of this bilocal field the averaged partition function is
\begin{align}
\begin{split}
    \overline{Z}=\int \mathcal{D}\chi_L\mathcal{D}\chi_R\exp\Biggl[  &-\frac{1}{2}\sum_{a,b}\int_\mathcal{C} dt_1~dt_2 \sum_j\chi_a^j(t_1)\Big[(\delta_{ab}\partial_{t_1}-\mu_{ab}(t_1))\delta(t_1-t_2)\Big]\chi_b^j(t_2)\\
    &-\frac{N}{2}\frac{J^4}{4}\sum_{a,b}\int_\mathcal{C} dt_1 dt_2f_a(t_1)f_b(t_2)\left[O_{ab}(t_1,t_2)\right]^4\Biggr]~,
\end{split}
\end{align}
where the matrix $\mu_{ab}$ is
\begin{equation}
    \mu_{ab}(t)=\begin{pmatrix}
0 & \mu(t) \\
-\mu(t) & 0
\end{pmatrix}~.
\end{equation}
We can insert unity as
\begin{align}
\begin{split}
    1&=\int \mathcal{D}G_{ab}~\delta\left[N\left(G_{ab}(t_1,t_2)-O_{ab}(t_1,t_2)\right)\right]\\
    &\sim \int \mathcal{D}G_{ab}\mathcal{D}\Sigma_{ab}\exp\left[-\frac{N}{2}\int_\mathcal{C} dt_1 dt_2 \Sigma_{ab}(t_1,t_2)\left(G_{ab}(t_1,t_2)-O_{ab}(t_1,t_2)\right)\right]~.
\end{split}
\end{align}
Therefore, the averaged partition function is
\begin{align}
\begin{split}
    \overline{Z}=&\int\mathcal{D}\chi_L\mathcal{D}\chi_R\mathcal{D}G_{ab}\mathcal{D}\Sigma_{ab}\\
    &\times\exp\left[-\frac{1}{2}\sum_{a,b}\int_\mathcal{C} dt_1 dt_2\sum_j\chi_a^j(t_1)\Big[ (\delta_{ab}\partial_{t_1}-\mu_{ab}(t_1))\delta_{\mathcal{C}}(t_1-t_2)+i\Sigma_{ab}(t_1,t_2) \Big]\chi_b ^j(t_2)\right]\\
    &\times \exp\left[-\frac{N}{2}\int_\mathcal{C} dt_1dt_2\sum_{a,b}\Big(\Sigma_{ab}(t_1,t_2)G_{ab}(t_1,t_2)+\frac{J^2}{4}f_a(t_1)f_b(t_2)G_{ab}(t_1,t_2)^4\Big)\right]~.
    \label{eq:AvgZrealt}
\end{split}
\end{align}
Now we want to integrate over the fermions. For that, let's define $\left[G_0^{-1}\right]_{ab}\equiv i\delta_{ab}\partial_{t_1}\delta_\mathcal{C}(t_1-t_2)$. Then, the integral to do is
\begin{align}
\begin{split}
    \int\mathcal{D}\chi_L\mathcal{D}\chi_R&\exp\left[-\frac{1}{2}\sum_{a,b}\sum_j\int_\mathcal{C}dt_1dt_2\chi_a^j(t_1)\Big[-i\left[G_0^{-1}\right]_{ab}-\mu_{ab}(t_1)\delta(t_1-t_2)+i\Sigma_{ab}(t_1,t_2)\Big]\chi_b^j(t_2) \right]\\
    &=\Big[\det\left(-i\left[G_0^{-1}\right]_{ab}(t_1,t_2)-\mu_{ab}(t_1)\delta(t_1-t_2)+i\Sigma_{ab}(t_1,t_2)\right)\Big]^{N/2}\\
    &=\exp\left[\frac{N}{2}\log\det\left(-i\left[G_0^{-1}\right]_{ab}(t_1,t_2)-\mu_{ab}(t_1)\delta(t_1-t_2)+i\Sigma_{ab}(t_1,t_2)\right)\right]~.
\end{split}
\end{align}
Therefore, the averaged partition function can be written in terms of an effective action $S_{eff}[G,\Sigma]$,
\begin{equation}
    \overline{Z}=\int\mathcal{D}G_{ab}\mathcal{D}\Sigma_{ab}~e^{iS_{eff}[G,\Sigma]}~,
\end{equation}
with
\begin{align}
\begin{split}
    \frac{1}{N}iS_{eff}[G,\Sigma]=&\frac{1}{2}\log\det \left(-i\left[G_0^{-1}\right]_{ab}(t_1,t_2)-\mu_{ab}(t_1)\delta(t_1-t_2)+i\Sigma_{ab}(t_1,t_2)\right)\\
    &-\frac{1}{2}\int_\mathcal{C}dt_1dt_2\sum_{a,b}\Big(\Sigma_{ab}(t_1,t_2)G_{ab}(t_1,t_2)+\frac{J_L(t)J_R(t)}{4}G_{ab}(t_1,t_2)^4\Big)~,
\end{split}
\end{align}
where $J_a(t)\equiv J f_a(t)$. The saddle point equations are obtained as $\frac{\delta S_{eff}}{\delta G_{ab}}=0$ and $\frac{\delta S_{eff}}{\delta \Sigma_{ab}}=0$. The first variation immediately gives
\begin{equation}
    \Sigma_{ab}(t_1,t_2)=-J_a(t_1)J_b(t_2)G_{ab}(t_1,t_2)^3~.
\end{equation}
For the other variation, we use that
\begin{equation}
    \delta(\log\det A)=\Tr\left(A^{-1}\delta A\right)~,
\end{equation}
where the trace is taken in both "$ab$" space and $t_1,t_2$ space. The equation $\frac{\delta S_{eff}}{\delta \Sigma_{ab}}=0$ implies that
\begin{equation}
    \left(G_0^{-1}(t_1,t_2)-i\mu(t_1)\delta(t_1-t_2)-\Sigma(t_1,t_2)\right)^{-1}_{ab}=G_{ab}(t_1,t_2)~.
\end{equation}
After inverting,
\begin{equation}
    \left[G_0^{-1}(t_1,t_2)\right]_{ab}-i\mu_{ab}(t_1)\delta(t_1-t_2)-\Sigma_{ab}(t_1,t_2)=\left[G^{-1}(t_1,t_2)\right]_{ab}~.
\label{eq:Dysoninv}
\end{equation}

In order to get the Kadanoff-Baym equations, we convolute from the right and from the left with $G_{ab}$, respectively, to obtain
\begin{align}
\begin{split}
    \int_\mathcal{C}dt\left[G_0^{-1}(t_1,t)\right]_{ac}G_{cb}(t,t_2)-i\mu_{ac}(t_1)G_{cb}(t_1,t_2)-\int_\mathcal{C}dt\Sigma_{ac}(t_1,t)G_{cb}(t,t_2)&=\delta_\mathcal{C}(t_1-t_2)\\
    \int_\mathcal{C}dt~G_{ac}(t_1,t)\left[G_0^{-1}(t,t_2)\right]_{cb}-i G_{ac}(t_1,t_2)\mu_{cb}(t_2)-\int_\mathcal{C}dt~G_{ac}(t_1,t)\Sigma_{cb}(t,t_2)&=\delta_\mathcal{C}(t_1-t_2)~.
\end{split}
\end{align}

In order to rephrase these equations in the Kadanoff-Baym form, we must replace countour integrals by time integrals along both branches $+$ and $-$ of the Schwinger-Keldysh contour. 
Now $t_1$ and $t_2$ can be on any of the two branches. It is convenient to define the greater and lesser Green's functions as
\begin{align}
\begin{split}
    G_{ab}^>(t_1,t_2)&=G_{ab}(t_1^-,t_2^+)\\
    G_{ab}^<(t_1,t_2)&=G_{ab}(t_1^+,t_2^-)~,
\end{split}
\end{align}
where $t_i^-$ lives on the lower contour, and $t_i^+$ lives on the upper contour. More explicitly, since $t_i^-$ is always later than $t_j^+$ in ${\cal C}$
\begin{align}
\begin{split}
    G_{ab}^>(t_1,t_2)&=-\frac{i}{N}\sum_j\langle\chi_a^j(t_1)\chi_b^j(t_2)\rangle\\
    G_{ab}^<(t_1,t_2)&=-\frac{i}{N}\sum_j\langle\chi_b^j(t_2)\chi_a^j(t_1)\rangle~.
\end{split}
\end{align}

From the definitions we have the relation $~G_{ab}^>(t_1,t_2)=-G_{ba}^<(t_2,t_1)$. If we want to get the greater component of $G_{ab}$ we can take $t_1=t_1^-$ and $t_2=t_2^+$. Then, using the expression for $\left[G_0^{-1}(t_1,t)\right]_{ab}$, we get
\begin{align}
\begin{split}
    i\partial_{t_1}G_{ab}^>(t_1,t_2)&=i\mu_{ac}(t_1)G_{cb}^>(t_1,t_2)+\int_\mathcal{C}dt \Sigma_{ac}(t_1,t)G_{cb}(t,t_2)\\
    -i\partial_{t_2}G_{ab}^>(t_1,t_2)&=iG_{ac}^>(t_1,t_2)\mu_{cb}(t_2)+\int_\mathcal{C}dt ~G_{ac}(t_1,t) \Sigma_{cb}(t,t_2)~.
\label{eq:KBcontour}
\end{split}
\end{align}
One can define the retarded, advanced, and Keldysh Green's functions as
\begin{align}
\begin{split}
    G_{ab}^R(t_1,t_2)&=\theta(t_1-t_2)\left[G_{ab}^>(t_1,t_2)-G_{ab}^<(t_1,t_2)\right]\\
    G_{ab}^A(t_1,t_2)&=-\theta(t_2-t_1)\left[G_{ab}^>(t_1,t_2)-G_{ab}^<(t_1,t_2)\right]\\
    G_{ab}^K(t_1,t_2)&=G_{ab}^>(t_1,t_2)+G_{ab}^<(t_1,t_2)~,
    \label{eq:retadvkel}
\end{split}
\end{align}
and similarly for the self energies,
\begin{align}
\begin{split}
    \Sigma_{ab}^R(t_1,t_2)&=\theta(t_1-t_2)\left[\Sigma_{ab}^>(t_1,t_2)-\Sigma_{ab}^<(t_1,t_2)\right]\\
    \Sigma_{ab}^A(t_1,t_2)&=-\theta(t_2-t_1)\left[\Sigma_{ab}^>(t_1,t_2)-\Sigma_{ab}^<(t_1,t_2)\right]\\
    \Sigma_{ab}^K(t_1,t_2)&=\Sigma_{ab}^>(t_1,t_2)+\Sigma_{ab}^<(t_1,t_2)~.
\end{split}
\end{align}

Then, applying the Langreth rules to rewrite the integral in terms of a single time variable, equation \eqref{eq:KBcontour} becomes
\begin{align}
\begin{split}
    i\partial_{t_1}G_{ab}^>(t_1,t_2)&=i\mu_{ac}(t_1)G_{cb}^>(t_1,t_2)+\int_{-\infty}^{\infty}dt~ \Sigma_{ac}^R(t_1,t)G_{cb}^>(t,t_2)+\int_{-\infty}^{\infty}dt~ \Sigma_{ac}^>(t_1,t)G_{cb}^A(t,t_2)\\
    -i\partial_{t_2}G_{ab}^>(t_1,t_2)&=i G_{ac}^>(t_1,t_2)\mu_{cb}(t_2)+\int_{-\infty}^{\infty}dt~ G_{ac}^R(t_1,t)\Sigma_{cb}^>(t,t_2)+\int_{-\infty}^{\infty}dt ~G_{ac}^>(t_1,t)\Sigma_{cb}^A(t,t_2)~.
\label{eq:KBeqsapp}
\end{split}
\end{align}

\section{Equilibrium initial conditions}\label{App:initialconds}
In order to obtain the initial equilibrium Green's functions at an inverse temperature $\beta$ it is useful to begin with the imaginary time formulation of the model, where the hamiltonian takes the form as in \cite{MaldaQi}
\begin{equation}   
    H=\sum_{a=L,R}\frac{1}{4!}\sum_{ijkl}J_{ijkl}\chi_a^i\chi_a^j\chi_a^k\chi_a^l+i\mu\sum_j \chi_L^j\chi_R^j~.
\end{equation}

Following analogous steps as in the real-time case (see App. \ref{App:KBequations} for details), the imaginary-time averaged partition function can be written in terms of the time-ordered correlators at the saddle point,
\begin{equation}
    G_{ab}(\tau,\tau')=\frac{1}{N}\sum_j\langle \chi_a^j(\tau)\chi_b^j(\tau')\rangle~,
\end{equation}
and it is given by
\begin{align}
\begin{split}
    -\frac{S_{eff}[G,\Sigma]}{N}=&~\frac{1}{2}\log\det \left(\delta(\tau-\tau')\left(\delta_{ab}\partial_\tau-\mu \sigma_{ab}^y\right)-\Sigma_{ab}(\tau,\tau')\right)\\
    &-\frac{1}{2}\int d\tau d\tau'\sum_{a,b}\Sigma_{ab}(\tau,\tau')G_{ab}(\tau,\tau')+\frac{1}{2}\frac{J^2}{4}\sum_{a,b}\int d\tau d\tau' \left[G_{ab}(\tau,\tau')\right]^4~.
\label{eq:Seffimag}
\end{split}
\end{align}

Again, the saddle-point equations are obtained from $\delta S_{eff}/\delta \Sigma_{ab}=0$ and $\delta S_{eff}/\delta G_{ab}=0$. The variation $\delta S_{eff}/\delta G_{ab}=0$ gives immediately
\begin{equation}
    \Sigma_{ab}(\tau)=J^2 G_{ab}(\tau)^3~.    
    \label{eq:Sigmaapp}
\end{equation}
It is useful to do the variation $\delta S_{eff}/\delta \Sigma_{ab}$ in Matsubara frequency space, using the convention
\begin{equation}
    f(\tau)=\frac{1}{\beta}\sum_n e^{-i\omega_n \tau}f(i\omega_n)~, ~~~~~~f(i\omega_n)=\int_{0}^{\beta}d\tau e^{i\omega_n \tau}f(\tau)~,
\label{eq:Matsubara}
\end{equation}
where $\omega_n=(2n+1)\pi/\beta$ is the $n^{th}$ Matsubara frequency. The frequency-domain counterpart of \eqref{eq:Seffimag} is
\begin{align}
\begin{split}
    \frac{S_{eff}[G,\Sigma]}{N}=&-\frac{1}{2}\sum_n\log\det \begin{pmatrix}
i\omega_n+\Sigma_{LL}(i\omega_n) & i\mu-\Sigma_{LR}(i\omega_n) \\
-i\mu+\Sigma_{LR}(i\omega_n) & i\omega_n+\Sigma_{LL}(i\omega_n)
\end{pmatrix}
\\
&+\frac{1}{2}\sum_{a,b}\left[-\sum_n\Sigma_{ab}(i\omega_n)G_{ba}(i\omega_n)-\frac{J^2}{4}\int d\tau d\tau' \left[G_{ab}(\tau,\tau')\right]^4\right]
\label{eq:SeffimagMatsu}
\end{split}
\end{align}
and the equation $\delta S_{eff}/\delta \Sigma_{ab}=0$ gives the Dyson equations
\begin{align}
\begin{split}
    G_{LL}(i\omega_n)&=-\frac{i\omega_n+\Sigma_{LL}}{(i\omega_n+\Sigma_{LL})^2+(i\mu-\Sigma_{LR})^2}\\[10pt]
    G_{LR}(i\omega_n)&=-\frac{i\mu-\Sigma_{LR}}{(i\omega_n+\Sigma_{LL})^2+(i\mu-\Sigma_{LR})^2}~,
\label{eq:SDeqsApp}
\end{split}
\end{align}
with the self-energies given by \eqref{eq:Sigmaapp}. These are the equations that NESSi uses to determine the equilibrium initial condition.

However, for the implementation of the "standard" predictor-corrector method (which we use in the simpler cases to compare with NESSi's results), the equilibrium solution is found by the analytically continued version of these equations, following the proposals in \cite{Banerjee_2017,Plugge_2020,Sahoo_2020,Haenel_2021}. The two Dyson equations \eqref{eq:SDeqsApp} are analitically continued by the standard procedure $i\omega_n\rightarrow \omega +i\delta$, which connects the Matsubara and retarded propagators:
\begin{align}
\begin{split}
    G_{LL}^R(\omega)&=-\frac{\omega+i\delta +\Sigma_{LL}}{(\omega+i\delta +\Sigma_{LL})^2+(i\mu-\Sigma_{LR})^2}\\[10pt]
    G_{LR}^R(\omega)&=-\frac{i\mu-\Sigma_{LR}}{(\omega+i\delta +\Sigma_{LL})^2+(i\mu-\Sigma_{LR})^2}~.
\label{eq:SDeqsrealApp}
\end{split}
\end{align}
The equation for $\Sigma$ is written in frequency space before performing the analytic continuation:
\begin{equation}
    \Sigma_{ab}(i\omega_n)=\frac{J^2}{\beta^2}\sum_{n_1,n_2}G_{ab}(i\omega_{n_1})G_{ab}(i\omega_{n_2})G_{ab}(i\omega_n-i\omega_{n_1}-i\omega_{n_2})~.
    \label{eq:SigmaomegaApp}
\end{equation}
If we express the Green's functions through their spectral representation,
\begin{equation}
    G_{LL}(i\omega_k)=-\int d\omega \frac{\rho_{LL}(\omega)}{i\omega_k-\omega},~~~~iG_{LR}(i\omega_k)=\int d\omega \frac{\rho_{LR}(\omega)}{i\omega_k-\omega}~,
\end{equation}
where
\begin{equation}
    \rho_{LL}(\omega)=\frac{1}{\pi}\Im G_{LL}^R(\omega),~~~~\rho_{LR}(\omega)=-\frac{1}{\pi}\Im[iG_{LR}^R(\omega)]
\label{eq:spectdensApp}
\end{equation}
are the spectral functions, eq. \eqref{eq:SigmaomegaApp} becomes 
\begin{equation}
    \Sigma_{ab}(i\omega_n)=-\tilde{\sigma}_{ab}\frac{J^2}{\beta^2}\int \prod_{i=1}^{3}\left(d\omega_i\rho_{ab}(\omega_i)\right)\sum_{n_1}\frac{1}{i\omega_{n_1}-\omega_1}\sum_{n_2}\frac{1}{i\omega_{n_2}-\omega_2}\frac{1}{i\omega_{n_2}-\Omega}~,
\end{equation}
with $\tilde{\sigma}_{LL}=-1$, $\tilde{\sigma}_{LR}=i$ and $\Omega=i\omega_n-i\omega_{n_1}-\omega_3$. The two Matsubara sums over frequencies can be evaluated\footnote{For instance, the sum over $n_2$ gives $\frac{1}{\beta}\sum\limits_{n_2}\frac{1}{i\omega_{n_2}-\omega_2}\frac{1}{i\omega_{n_2}-\Omega}=\frac{n_F(\omega_2)-n_F(\Omega)}{\omega_2-\Omega}$, where $n_F(\omega)=\frac{1}{e^{\beta \omega}+1}$ is the Fermi distribution function. By using the symmetries of the Fermi distribution function the remaining sum can be simplified and computed as $\sum_{n_1}\frac{1}{i\omega_{n_1}-\omega_1}\frac{1}{i\omega_{n_1}-\tilde{\Omega}}=-\frac{n_F(\omega_1)-n_F(\tilde{\Omega})}{i\omega_n-\omega_1-\omega_2-\omega_3}$, with $\tilde{\Omega}=i\omega_n-\omega_2-\omega_3$}, and after some manipulations using the Fermi/Bose distribution functions, we obtain
\begin{equation}
    \Sigma_{ab}(i\omega_n)=\tilde{\sigma}_{ab}J^2\int \prod_{i=1}^{3}\left(d\omega_i\rho_{ab}(\omega_i)\right)\frac{\left[n_F(\omega_1)n_F(\omega_2)n_F(\omega_3)+n_F(-\omega_1)n_F(-\omega_2)n_F(-\omega_3)\right]}{i\omega_n-\omega_1-\omega_2-\omega_3}~.
\label{eq:SigmaiomeganApp}
\end{equation}
Now we can analitiycally continue, $i\omega_n\rightarrow \omega+i\delta$, and obtain the retarded self-energy $\Sigma_{ab}^R(\omega)$:
\begin{equation}
    \Sigma_{ab}^R(\omega)=\tilde{\sigma}_{ab}J^2\int \prod_{i=1}^{3}\left(d\omega_i\rho_{ab}(\omega_i)\right)\frac{\left[n_F(\omega_1)n_F(\omega_2)n_F(\omega_3)+n_F(-\omega_1)n_F(-\omega_2)n_F(-\omega_3)\right]}{\omega-\omega_1-\omega_2-\omega_3+i\delta}~.
\label{eq:SigmacontApp}
\end{equation}
Using the identity $\frac{1}{\Omega+i\eta}=-i\int_{0}^{\infty}dt e^{i(\Omega+i\eta)t}$, with $\Omega=\omega-\omega_1-\omega_2-\omega_3$, and defining the "time-dependent occupations"
\begin{equation}
    n_{ab}^s(t)\equiv\int_{-\infty}^{\infty}d\omega \rho_{ab}(\omega)n_F(s\omega)e^{-i\omega t}
\end{equation}
we can write
\begin{equation}
    \Sigma_{ab}^R(\omega)=-i\tilde{\sigma}_{ab}J^2\int_{0}^{\infty}dt e^{i(\omega+i\eta)t}\left[n_{ab}^+(t)^3+n_{ab}^-(t)^3\right]~.
\end{equation}
Finally, noticing that 
\begin{equation}
    n_{ab}^-(t)=\sigma_{ab}\left[n_{ab}^+(t)\right]^*~,
\end{equation}
where $\sigma_{LL}=1$, $\sigma_{LR}=-1$, we can supress the $s=\pm$ label and work with a single function
\begin{equation}
    n_{ab}(t)\equiv\int_{-\infty}^{\infty}d\omega\rho_{ab}(\omega)n_F(\omega)e^{-i\omega t}~.
\label{eq:occupationsApp}
\end{equation}
Then, 
\begin{align}
\begin{split}
    \Sigma_{LL}^R(\omega)&\sim n_{LL}^3(t)+\left[n_{LL}^3(t)\right]^*=2\Re\left[n_{LL}^3(t)\right]\\
    \Sigma_{LR}^R(\omega)&\sim n_{LL}^3(t)-\left[n_{LL}^3(t)\right]^*=2i\Im\left[n_{LR}^3(t)\right]
\end{split}
\end{align}
and we can write explicitly our final expressions for the self-energies:
\begin{align}
\begin{split}
    \Sigma_{LL}^R(\omega)&=2iJ^2\int_{0}^{\infty}dt e^{i(\omega+i\delta)t}\Re\left[n_{LL}^3(t)\right]\\
    \Sigma_{LR}^R(\omega)&=2iJ^2\int_{0}^{\infty}dt e^{i(\omega+i\delta)t}\Im\left[n_{LR}^3(t)\right]~.
\label{eq:SigmaReImApp}
\end{split}
\end{align}

With this, the numerical procedure to find an equilibrium solution consists in the following steps: we start with an initial guess for $G_{LL}(\omega)$, $G_{LR}(\omega)$, which we take to be the free ones (i.e., the ones for $J=0$), which from the Schwinger-Dyson equations \eqref{eq:SDeqsrealApp} are given by
\begin{align}
\begin{split}
    G_{LL}^{free}&=-\frac{\omega+i\delta}{(\omega+i\delta)^2-\mu^2}\\[10pt]
    G_{LR}^{free}&=-\frac{i\mu}{(\omega+i\delta)^2-\mu^2}~.
\label{eq:freeGFsApp}
\end{split}
\end{align}

From these we compute the spectral densities $\rho_{LL}(\omega)$ and $\rho_{LR}(\omega)$ by virtue of \eqref{eq:spectdensApp}, and we use them to compute the "time-dependent occupations" \eqref{eq:occupationsApp}, which we then plug into \eqref{eq:SigmaReImApp} to get the retarded self-energies $\Sigma_{LL}^R(\omega)$ and $\Sigma_{LR}^R(\omega)$. Finally we plug them in the SD equations \eqref{eq:SDeqsrealApp} and obtain new functions $G_{LL}^R(\omega)$, $G_{LR}^R(\omega)$. We repeat this procedure until the solutions converge to the desired accuracy.

\section{Time-dependent energy}\label{App:energies}
From the time evolution equation of a Majorana field,
\begin{equation}
    \partial_t\chi_a^j=i\left[H,\chi_a^j\right]~,
\end{equation}
we can compute the time derivative of $G_{LL}^>(t_1,t_2)$. As
\begin{equation}
    G_{LL}^>(t_1,t_2)=-\frac{i}{N}\sum_j\langle\chi_L^j(t_1)\chi_L^j(t_2)\rangle~,
\end{equation}
we have
\begin{equation}
    \partial_{t_1}G_{LL}^>(t_1,t_2)=-\frac{i}{N}\sum_j\langle\partial_{t_1}\chi_L^j(t_1)\chi_L^j(t_2)\rangle=\frac{1}{N}\sum_j\langle\left[H,\chi_L^j(t_1)\right]\chi_L^j(t_2)\rangle~.
\end{equation}
Let's take $t_1=t_2=t$ and use
\begin{equation}
    \sum_j\left[H,\chi_L^j(t)\right]\chi_L^j(t)=4H_L(t)+H_{int}(t)~,
\end{equation}
which can be proved by explicitly computing the commutator. Then,
\begin{equation}
    N\partial_tG_{LL}^>(t,t)=4\langle H_L(t)\rangle +\langle H_{int}(t)\rangle~,
\end{equation}
where $\langle H_{int}\rangle= i\mu(t)\sum_j\langle\chi_L^j(t)\chi_R^j(t)\rangle=-\mu(t) NG_{LR}^>(t,t)$. Then, the energy $E_L\equiv \langle H_L(t)\rangle$ is given by
\begin{equation}
    \frac{1}{N}E_L(t)=\frac{1}{4}\partial_t G_{LL}^>(t,t)+\frac{\mu(t)}{4}G_{LR}^>(t,t)~.
\label{eq:expHt}
\end{equation}

We can use the Kadanoff-Baym equations \eqref{eq:KBeqsapp} to remove the time derivative from this expression as
\begin{equation}
    i\partial_{t_1}G_{LL}^>(t_1,t_2)=i\mu(t_1)G_{RL}^>(t_1,t_2)+\int_{-\infty}^{t_1}dt'~ \Sigma_{Lc}^R(t_1,t')G_{cL}^>(t',t_2)+\int_{-\infty}^{t_2}dt'~ \Sigma_{Lc}^>(t_1,t')G_{cL}^A(t',t_2)~,
\end{equation}
where $c$ is summed over $L,R$. We can write each integral as
\begin{align}
    \int_{-\infty}^{t_1}dt'~ \Sigma_{Lc}^R(t_1,t')G_{cL}^>(t',t_2)&=\int_{-\infty}^{t_1}dt'~\left[\Sigma_{Lc}^>(t_1,t')-\Sigma_{Lc}^<(t_1,t')\right]G_{cL}^>(t',t_2)\\
    \int_{-\infty}^{t_2}dt'~ \Sigma_{Lc}^>(t_1,t')G_{cL}^A(t',t_2)&=\int_{-\infty}^{t_2}dt~\Sigma_{Lc}^>(t_1,t')\left[G_{cL}^<(t',t_2)-G_{cL}^>(t',t_2)\right]~.
\end{align}

We can now split the second integral as $\int_{-\infty}^{t_2}\rightarrow \int_{-\infty}^{t_1}+\int_{t_1}^{t_2}$. We can combine $\int_{-\infty}^{t_1}$ with the first integral, and forget about the integral $\int_{t_1}^{t_2}$, since this term will vanish when $t_1=t_2$ (recall \eqref{eq:expHt}).
Combining everything together and expanding the index $c=L,R$, we obtain
\begin{align}
\begin{split}
    i\partial_{t_1}G_{LL}^>(t_1,t_2)=i\mu(t_1)G_{RL}^>(t_1,t_2)+\int_{-\infty}^{t_1} dt'\Bigl[&\Sigma_{LL}^>(t_1,t')G_{LL}^<(t',t_2)-\Sigma_{LL}^<(t_1,t')G_{LL}^>(t',t_2)\\
    +&\Sigma_{LR}^>(t_1,t')G_{RL}^<(t',t_2)-\Sigma_{LR}^<(t_1,t')G_{RL}^>(t',t_2)\Bigr]+...~,
\end{split}
\end{align}
where the dots refer to the terms that will vanish when $t_1=t_2$. We can finally write the self energies in terms of the Green's functions according to \eqref{eq:Sigmarealt}, use the symmetry $G_{RL}^<(t_1,t_2)=-G_{LR}^>(t_2,t_1)$ and set $t_1=t_2$ to arrive to a final expression for the energy
\begin{equation}
    \frac{1}{N}E_L(t)=-\frac{iJ_L(t)}{4}\int_{-\infty}^{t}dt'\Big[J_L(t')\left(G_{LL}^>(t,t')^4-G_{LL}^<(t,t')^4\right)+J_R(t')\left(G_{LR}^>(t,t')^4-G_{LR}^<(t,t')^4\right)\Big]~.
\end{equation}

\bibliographystyle{JHEP}
\bibliography{SYKpaper}

\end{document}